\RequirePackage{fix-cm}
\RequirePackage{amsmath}

\documentclass[smallcondensed, final]{svjour3} 

\smartqed  

\usepackage{lineno,hyperref}
\modulolinenumbers[5]
\usepackage{multirow}
\usepackage{caption}
\usepackage{subcaption}
\usepackage{graphicx}
\usepackage{epstopdf}
\usepackage{color}
\usepackage{amsmath}
\usepackage{wrapfig}

\journalname{Knowledge and Information Systems}

\usepackage{algorithm2e}
\SetAlCapSkip{1em}
\newcommand{\asgn}{\mathrel{=}}

\newcommand{\WT}{{\it wtree}}
\newcommand{\PWT}{{\tt pwt}}
\newcommand{\DD}{{\tt dd}}
\newcommand{\levelWT}{{\tt levelWT}}
\newcommand{\sortWT}{{\tt sortWT}}
\newcommand{\access}[2]{\mathit{access}(#1,#2)}
\newcommand{\rank}[3]{\mathit{rank}_{#1}(#2,#3)}
\newcommand{\select}[3]{\mathit{select}_{#1}(#2,#3)}

\DeclareMathSymbol{\mlq}{\mathrel}{operators}{``}
\DeclareMathSymbol{\mrq}{\mathrel}{operators}{`'}

\begin{document}

\title{Parallel Construction of Wavelet Trees on Multicore
  Architectures\thanks{A previous version of this paper appeared in the 13th
    International Symposium on Experimental Algorithms (SEA
    2014)\cite{Fuentes2014}.}
  \thanks{This work was supported in part by the European Union's Horizon 2020
research and innovation programme under the Marie Sk{\l}odowska-Curie grant
agreement No 690941 and the doctoral scholarships of CONICYT No 21120974 and
63130228 (first and second authors, respectively). We also would like to thank
Roberto As\'in for making his multicore computers, Mastropiero and G\"unther
Frager, available to us.}
  \thanks{This is an Author's Original Manuscript of an article whose final and definitive form,
the Version of Record, has been published in Knowledge and Information Systems [copyright Springer],
available online at: \url{http://dx.doi.org/10.1007/s10115-016-1000-6}.}
}


\author{Jos\'e Fuentes-Sep\'ulveda \and Erick Elejalde \and Leo Ferres
  \and Diego Seco }

\authorrunning{J. Fuentes-Sep\'ulveda, E. Elejalde, L. Ferres,
  D. Seco} 

\institute{Jos\'e Fuentes-Sep\'ulveda \and Erick Elejalde \and Diego
  Seco \at Department of Computer Science, Universidad de
  Concepci\'on, Chile.\\ \email{\{jfuentess,eelejalde,dseco\}@udec.cl}
  \and Leo
  Ferres \at Faculty of Engineering, Universidad del Desarrollo, Chile \\
  \email{lferres@udd.cl }
  \and
  Corresponding authors: Jos\'e Fuentes-Sep\'ulveda \and Leo Ferres}

\date{}

\maketitle

\begin{abstract}
The wavelet tree has become a very useful data structure to
efficiently represent and query large volumes of data in many
different domains, from bioinformatics to geographic information
systems. One problem with wavelet trees is their construction time. In
this paper, we introduce two algorithms that reduce the time
complexity of a wavelet tree's construction by taking advantage of
nowadays ubiquitous multicore machines.

Our first algorithm constructs all the levels of the wavelet in
parallel in $O(n)$ time and $O(n\lg\sigma + \sigma\lg n)$ bits of working space,
where $n$ is the size of the input sequence and $\sigma$ is the size of the
alphabet. Our second algorithm constructs the wavelet tree in a
domain-decomposition fashion, using our first algorithm in each
segment, reaching $O(\lg n)$ time and $O(n\lg\sigma + p\sigma\lg n/\lg\sigma)$
bits of extra space, where $p$ is the number of available cores. Both algorithms
are practical and report good speedup for large real datasets.


\end{abstract}

\section{Introduction and motivation}
\label{sec:intro}
After their introduction in the mid-2000s, {\em multicore} computers
--computers with more than one processing unit (called {\em cores})
and shared memory-- have become pervasive. In fact, nowadays it is
difficult to find a single-processor desktop, let alone a high-end
server. The argument in favor of multicore systems is simple:
thermodynamic and material considerations prevent chip manu\-facturers
from ever increasing clock rates. Since 2005, clock frequencies have
stagnated at around 3.75GHz for commodity computers, and even in 2015,
4GHz computers are still high-end (the Intel Core i7-4790K is a prime
example). Thus, the next step in performance is to take advantage of
the processing power of multicore computers. In order to do this,
algorithms and data structures will have to be modified to make them
behave well in these parallel architectures.

At the same time, the amount of data to be processed has become large
enough that the ability to maintain it close to the processor is
vital. This, in turn, has generated a keen interest in succinct data
structures, which besides reducing storage requirements, may also reduce the number of memory transfers,
the energy consumption, and can be used in low-capacity devices such as
smartphones. One such structure that has benefited from thorough
research is the {\em wavelet tree} \cite{bib:grossi03}, henceforth
\WT{}.  Although the \WT{} was originally devised as a data
structure for encoding a reordering of the elements of a sequence
\cite{bib:grossi03,bib:fmmn07}, it has now been successfully used in
many critical applications such as indexing documents~\cite{bib:vm07},
in processing grids~\cite{NNRtcs13} and sets of
rectangles~\cite{BLNSis13}, to name but a few. Two excellent
surveys have been written about this data structure
\cite{Navarro2012,Makris12}, and we refer the readers to them for more
application examples and details.

These succinct data structures, however, are generally quite expensive
(in time) to build, particularly as the size of the alphabet and the
size of the input increase, as is the case nowadays with the so-called
``big data'' revolution. We believe parallel computing is a good tool
for speeding up the processing of succinct data
structures. Unfortunately, (practical) parallel computing suffers from
several drawbacks that make these high-performant algorithms difficult
to come by: maintaing thread independence while communicating results,
keeping clear of ``thrashing'' the memory hierarchy are two such
problems.  Thus, a sizeable contribution to the state-of-the-art would
involve designing algorithms with good theoretical running times
that are also practical in modern commodity architectures with more
than one core, which would also help speed up processing of the target
data structures in distributed systems: if one node is faster, the
whole cluster would be faster as well.

\paragraph{Motivating example.}
Perhaps one of the prominent areas of research in the last
few years has been the analysis of genomic data
\cite{Schnattinger201213,Singer2012,NVBIO}. In combination with the
{\em Burrows-Wheeler transform} \cite{Burrows94ablock-sorting}, the
\WT{} has been used to construct compressed full-text indexes (the
{\em FM-index} \cite{Ferragina:2000:ODS:795666.796543,Ferragina2004})
over DNA sequences. The structure supports efficient algorithms for
important problems in bioinformatics such as the identification of
patterns (like the mutations that are known to cause some diseases) or
the alignment of short DNA sequences, known as reads (which is a
fundamental step to reconstruct a genome), all this {\em without
decompressing the data}. The cost of DNA sequencing has plummeted in
the last few years thanks to next-generation sequencing technologies
\cite{DNASequencingCosts}. In addition, these technologies are also
much faster. For example, in 2005, a single sequencing run could
generate at most one gigabase of data. Meanwhile, in 2014, a single sequencing
run could generate up to 1.8 terabases of data \cite{illumina}.
These two factors have drastically increased the amount
of genomic data to be processed. Therefore, full-text indexes based on
\WT{}s need to be updated periodically. These updates do not modify
the already indexed data but add new sequences. This process is not
trivial because the {\em Burrows-Wheeler transform} is a
reorganization of the whole sequence in order to make it more
compressible. In order to support these updates there are two options:
the use of fully dynamic \WT{}s or the periodic reconstruction of the
\WT{} (a solution used in other domains such as Web search
engines). Dynamic versions of \WT{}s are quite slow in both update and
rank/select operations (see Section 3.2 of \cite{Makris12}). The other
option is the usage of a static \WT{} and a buffer (which stores the
updates since the last reconstruction of the static index). To support
queries, both the static \WT{} and the buffer are used. When the
buffer is full, the static \WT{} is reconstructed considering the
symbols on the buffer, which is emptied after that. Thus, improving
the construction time of static \WT{}s becomes critical, for example,
to provide solutions in this kind of dynamic domain in which queries
are much more frequent than updates.

In this paper, we propose two parallel algorithms for the most
expensive operation on \WT{}s: its construction. The first algorithm,
\PWT{}, has $O(n)$ time complexity and uses $O(n\lg\sigma+\sigma\lg
n)$ bits of space\footnote{We use $\lg x = \log_2 x$.}, including the
space of the final \WT{} and excluding the input, where $\sigma$ is the
size of the alphabet. The second
algorithm, \DD{}, is an improved version of the \DD{} algorithm
presented on \cite{Fuentes2014}. This new version has $O(\lg n)$ time
complexity and uses $O(n\lg\sigma+p\sigma\lg n/\lg\sigma)$ bits of
space, using $p$ threads (see Sect. \ref{sec:pwt}). The \PWT{}
algorithm improves the $O(n)$ memory consumption in
\cite{Shun2015}. Meanwhile, the new \DD{} algorithm improves the
$O(n)$ time complexity of our previous work \cite{Fuentes2014} and the
time complexity of \cite{Shun2015} by a factor of $O(\lg\sigma)$. We
report experiments that demonstrate the algorithms to be not only
theoretically good, but also practical for large datasets on commodity
architectures, achieving good speedup (Sect. \ref{sec:exps}). As far
as we can tell, we use the largest datasets to-date and our algorithms are
faster for most use cases than the state-of-the-art \cite{Shun2015}.


\section{Background and related work}
\label{sec:relwork}

\subsection{Dynamic multithreading model}
\label{subsec:dynthreads}
{\em Dynamic multithreading} (DYM) \cite[Chapter 27]{Cormen2009} is a
model of parallel computation which is faithful to several industry
standards such as Intel's CilkPlus (\url{cilkplus.org}), OpenMP Tasks
(\url{openmp.org/wp}), and Threading Building Blocks
(\url{threadingbuildingblocks.org}). Besides its mathematical rigour,
it is precisely this adoption by many high-end compiler vendors that
make the model so appealing for practical parallel algorithms.

In the {\em Dynamic multithreading} model, a {\em multithreaded
  computation} is defined as a directed acyclic graph (DAG) $G=(V,E)$,
where the set of vertices $V$ are instructions and $(u,v) \in E$ are
dependencies between the instructions; whereby in this case, $u$ must
be executed before $v$.\footnote{Notice that the RAM model is a subset
  of the DYM model where the outdegree of every vertex $v \in V$ is
  $\leq 1$.} In order to signal parallel execution, we augment
sequential pseudocode with three keywords, {\bf spawn}, {\bf sync} and
{\bf parfor}. The {\bf spawn} keyword signals that the procedure call
that it precedes {\em may be} executed in parallel with the next
instruction in the instance that executes the {\bf spawn}. In turn,
the {\bf sync} keyword signals that all spawned procedures must finish
before proceeding with the next instruction in the stream. Finally,
{\bf parfor} is simply ``syntactic sugar'' for {\bf spawn}'ing and
{\bf sync}'ing ranges of a loop iteration. If a stream of instructions
does not contain one of the above keywords, or a {\bf return} (which
implicitly {\bf sync}'s) from a procedure, we group these instruction
into a single {\em strand}. The {\bf parfor} keyword, which
  we use repeatedly here, is implemented by halving the range of loop
  iterations, {\bf spawn}'ing one half and using the current procedure
  to process the other half recursively until reaching one iteration per
  range. After that, the iterations are executed in parallel. 
   This implementation adds an overhead to
  the parallel algorithm bounded above by the logarithm of the number
  of loop iterations.  For example, Algorithm \ref{fig:exampleAlgo}
  represents a parallel algorithm using {\bf parfor} and Figure
  \ref{fig:dym} shows its multithreaded computation. In the figure,
  each circle represents one strand and each rounded rectangle
  represents strands that belong to the same procedure call. The
  algorithm starts on the initial procedure call with the entire range
  $[0,7]$. The first half of the range is {\bf spawn}ed (black circle
  in the initial call) and the second half is processed by the same
  procedure (gray circle of the initial call). This divide-and-conquer
  strategy is repeated until reaching strands with one iteration of
  the loop (black circles on the bottom of the figure). Once an
  iteration is finished, the corresponding strand {\bf sync}s to its
  calling procedure (white circles), until reaching the final strand
  (white circle of the initial call). For more examples of the usage
  of the DYM model, see \cite[Chapter 27]{Cormen2009}. Strands are
scheduled onto cores using a {\em work-stealing} scheduler, which does
the load-balancing of the computations. Work-stealing schedulers have
been proved to be a factor of 2 away from optimal performance
\cite{Blumofe:1999:SMC:324133.324234}.

\noindent\begin{minipage}{.38\textwidth} \vspace{-1.5cm}
  \begin{algorithm}[H] \SetAlgoNoEnd \SetKwFor{PFor}{parfor}{do}{end}
  \DontPrintSemicolon
  $A:$ array of $8$ numbers\;

  \PFor{$i\asgn 0$ \KwTo $7$}{$A[i] = 0$\;}
  \Return
  \caption{Example of a parallel algorithm using the {\bf
        parfor} keyword. In parallel, the algorithm initializes all
      the elements of the array $A$ with $0$.}
   \label{fig:exampleAlgo}
  \end{algorithm}
 \end{minipage} \quad
\begin{minipage}[t]{.6\textwidth} \centering
    \includegraphics[scale=.33]{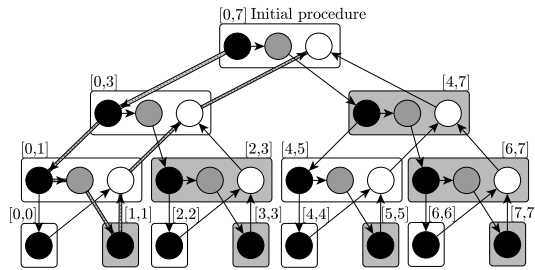}
    \captionof{figure}{Example of a multithreaded computation
on the Dynamic Multithreading Model. It corresponds to the Directed
Acyclic Graph representation of the Algorithm
\ref{fig:exampleAlgo}. Vertices represent strands and edges represent
dependences.}
   \label{fig:dym}
 \end{minipage}

 To measure the efficiency of our parallel wavelet tree algorithms, we
 use three metrics: the {\em work}, the {\em span} and {\em
   speedup}. In accordance to the parallel literature, we will
 subscript running times by $p$, so $T_p$ is the running time of an
 algorithm on $p$ cores. The {\em work} is the total running time
 taken by all strands when executing on a {\em single} core (i.e.,
 $T_1$), while the {\em span}, denoted as $T_\infty$, is the {\em
   critical path} (the longest path) of $G$. In Figure
   \ref{fig:dym}, assuming that each strand takes unit time, the work
   is $29$ time units and the span is $8$ time units (this is
   represented in the figure with thicker edges). In this paper, we
 are interested in speeding up wavelet tree manipulation and improving
 the lower bounds of this speedup. To measure this, we will define
 {\em speedup} as $T_1/T_P = O(p)$, where linear speedup
 $T_1/T_p = \Theta(p)$, is the goal and the theoretical upper
 bound. We also define {\em parallelism} as the ratio
 $T_1/T_{\infty}$, the theoretical maximum number of cores for which
 it is possible to achieve linear speedup.

\subsection{Wavelet trees}
\label{subsec:wt}
A wavelet tree (\WT{}) is a data structure that maintains a sequence
of $n$ symbols $S=s_1,s_2,\ldots,s_n$ over an alphabet
$\Sigma=[1..\sigma]$ under the following operations: $\access{S}{i}$,
which returns the symbol at position $i$ in $S$; $\rank{c}{S}{i}$,
which counts the times symbol $c$ appears up to position $i$ in $S$;
and $\select{c}{S}{j}$, which returns the position in $S$ of the
$j$-th appearance of symbol $c$.  Storage space of \WT{}s can be
bounded by different measures of the entropy of the underlying data,
thus enabling compression. In addition, they can be implemented
efficiently \cite{bib:claude08} and perform well in practice.

The \WT{} is a balanced binary tree. We identify the two children of a
node as left and right. Each node represents a range $R \subseteq [1,
\sigma]$ of the alphabet $\Sigma$, its left child represents a subset
$R_l$, which corresponds with the first half of $R$, and its right
child a subset $R_r$, which corresponds with the second half.  Every
node virtually represents a subsequence $S'$ of $S$ composed of
symbols whose value lies in $R$. This subsequence is stored as a
bitmap in which a $0$ bit means that position $i$ belongs to $R_l$ and
a $1$ bit means that it belongs to $R_r$.

At its simplest, a \WT{} requires $n \lceil \lg \sigma \rceil + o (n
\lg \sigma)$ bits for the data, plus $O(\sigma \lg n)$ bits to store
the topology of the tree (considering one pointer per node), and
supports aforementioned queries in $O(\lg \sigma)$ time by traversing
the tree using $O(1)$-time $\mathit{rank/select}$ operations on
bitmaps \cite{Raman2007}. A simple recursive construction algorithm
takes $O(n \lg \sigma)$ time.  As mentioned before, the space required
by the structure can be reduced: the data can be compressed and stored
in space bounded by its entropy (via compressed encodings of bitmaps
and modifications on the shape of the tree), and the $O(\sigma \lg n)$
bits of the topology can be removed, effectively using one pointer per
{\em level} of the tree~\cite{bib:claude08}, which is important for
large alphabets. We focus on construction using a pointer per level
because, even though it adds some running time costs, it is more
suitable for \emph{big data}. This notwithstanding, it is trivial to
apply the technique to the one-pointer-per-node construction case, and
our results can be readily extended to other encodings and tree
shapes.

\begin{figure}
  \centering
  \begin{subfigure}[b]{\textwidth}
     \centering
     \includegraphics[scale=.8]{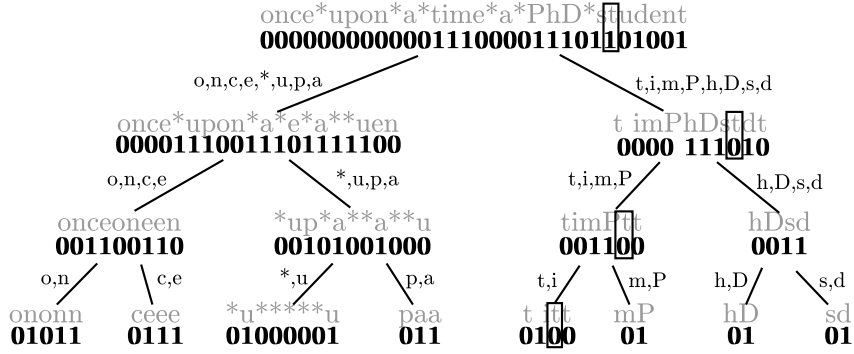}
     \caption{Representation of a \WT{} using one pointer per
         node and its associated bitmap. The subsequences of $S$ in
         the nodes (gray font) and the subsets of $\Sigma$ in the
         edges are drawn for illustration purposes.}
     \label{fig:wtree1}
  \end{subfigure}
  \begin{subfigure}[b]{\textwidth}
     \centering
     \includegraphics[scale=.8]{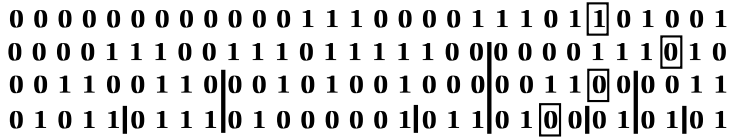}
     \caption{Representation of a \WT{} using one pointer per
         level and its associtaed $n$-bit bitmap. It can simulate the
         nagivation on the tree by using the rank operation over the
         bitmaps.}
     \label{fig:wtree2}
  \end{subfigure}
  \caption{A \WT{} for the sequence $S=$``{\tt once upon a
        time a PhD student}'' and the contiguous alphabet
      $\Sigma=\{${\tt o,n,c,e,` ',u,p,a,t,i,m,P,h,D,s,d}$\}$. We draw
      spaces using stars.}
  \label{fig:wtree}
\end{figure}

Figure \ref{fig:wtree} shows an example of two \WT{}
  representations for the sequence
  $S=``once~upon~a~time~a~PhD~student"$. Figure \ref{fig:wtree1} shows
  the one-pointer-per-node representation, while Figure
  \ref{fig:wtree2} shows the one-pointer-per-level representation. In
  our algorithms, we implemented the one-pointer-per-level
  representation; however, for clarity, we use the
  one-pointer-per-node representation to exemplify. In both
  representations we highlighted the traversal performed by the
  operation $\access{S}{24}$. To answer it, a top-down traversal of
  the \WT{} is performed: if a bit $0$ is found, we visit the left
  branch; if a $1$, the right branch is chosen. In the first
  representation, the query works as follows: Let $\mathit{curr}$ be
  the root, $B_{\mathit{curr}}$ be the bitmap of the current node,
  $i=24$ be the index of interest, $R$ be the range
  $[0,\sigma-1]=[0,15]$ and $\mathit{rank}_{c}(B_{\mathit{curr}},i)$
  be the number of $c$-bits up to position $i$ in
  $B_{\mathit{curr}}$. At the beginning, we inspect the bit
  $B_{\mathit{curr}}[i]$. Since the bit is $1$, we recompute
  $i=\mathit{rank}_{1}(B_{\mathit{curr}},i)-1=7$, change
  $\mathit{curr}$ to be the right child of $\mathit{curr}$ and halve
  $R=[8,15]$. Then, we repeat the process. Since
  $B_{\mathit{curr}}[i]=0$,
  $i=\mathit{rank}_{0}(B_{\mathit{curr}},i)-1=4$, $\mathit{curr}$ is
  updated to be the left child of $\mathit{curr}$ and $R=[8,11]$. Now,
  $B_{\mathit{curr}}[i]=0$,
  $i=\mathit{rank}_{0}(B_{\mathit{curr}},i)-1=2$, $\mathit{curr}$ is
  changed to be the left child of $\mathit{curr}$ and
  $R=[8,9]$. Finally, in the last level, $B_{\mathit{curr}}[i]=0$, so
  the range $R=[8,8]$ and the answer for $\access{S}{24}$ is
  $\Sigma[8]=\mlq t\mrq$.  $\rank{c}{S}{i}$ and $\select{c}{S}{i}$
  perform similar traversals to $\access{S}{i}$. For a more detailed
  explanation of \WT{} operations, see \cite{Navarro2012}. For the
  one-pointer-{\em per-level} representation, the procedure is
  similar, with the exception that the traversal of the tree must be simulated
  with rank operations over the bitmaps \cite{bib:claude08}.

Practical implementations of \WT{}s can be found in {\sc Libcds}
\cite{libcds} and {\sc Sdsl} \cite{sdsl}. {\sc Libcds} implements a
recursive construction algorithm that works by halving $\Sigma$ into
binary sub-trees whose left children are all $0$s and right children
are all $1$s, until $1$s and $0$s mean only one symbol in
$\Sigma$. {\sc Sdsl} implements an algorithm based in the idea of {\em
counting sort} that is more efficient in memory. The algorithm counts
the number of bits that will be placed in each node of the \WT{},
computing the position of each symbol in each level of the \WT{},
which avoids maintaining a permutation of the input. Both libraries
are the best current sequential implementations available, without
considering space efficient construction algorithms
\cite{bib:claude11,bib:tischler11}.

As of late, some work has been done in parallel processing of \WT{}s.
In~\cite{Arroyuelo12}, the authors explore the use of wavelet trees
in web search engines. They assume a distributed memory model and
propose partition techniques to balance the workload of processing
\WT{}s. Note that our work is complementary to theirs, as each node in
their distributed system can be assumed to be a multicore computer
that can benefit from our algorithms.  In~\cite{Ladra12}, the authors
explore the use of SIMD instructions to improve the performance of
\WT{}s and other string algorithms
\cite{DBLP:conf/spire/FaroK12}. This set of instructions can be
considered as low-level parallelism, since they use instructions in
modern processors that work by joining registers for some integer
computation, dealing with 128-bit integers at a time. We can also
benefit from their work as it may improve the performance of the
sequential parts of our algorithms. However, we leave this
optimization for future work.

In \cite{Fuentes2014}, we introduced the first two parallel
  algorithms for \WT~construction: \PWT~and \DD, both with $O(n)$ time
  complexity. The details of \PWT~and an improvement of \DD~are given
  on Sections \ref{subsec:pwt} and \ref{subsec:dd},
  respectively. Based on \cite{Fuentes2014}, Shun \cite{Shun2015}
  introduces two new parallel algorithms. The first algorithm, called
\levelWT{}, constructs the \WT{} level-by-level. In each of the
$\lceil \lg \sigma \rceil$ levels, the algorithm uses a parallel
prefix sum algorithm to compute the position of the bits, constructing
the nodes and their bitmaps in parallel with $O(n)$ work and
$O(\lg n)$ span, which results in $O(n\lg \sigma)$ work and
$O(\lg n\lg \sigma)$ span. The second algorithm, called \sortWT{},
constructs all levels in parallel, similar to our original \PWT{}, instead of
one-by-one.  For a level 
$l$, the \sortWT{} algorithm applies a parallel stable integer sorting
using the $l$ most significant bits of each symbol as the key. With
the sorted input sequence, the algorithm fills the corresponding
bitarrays in parallel, using parallel prefix sum and filter algorithms
to compute the position of the bits.  The total work of the \sortWT{}
algorithm is $O(W_{\mathit{sort}}\lg \sigma)$, where
$W_{\mathit{sort}}$ is the work incurred by sorting, and
$O(S_{\mathit{sort}}+\lg n)$ is the span, and where, in turn,
$S_{\mathit{sort}}$ corresponds to the span of the sorting algorithm
and the $\lg n$ component is the span of the prefix sum and filter
algorithms. The author also discusses a variation of the \sortWT{}
algorithm, reaching $O(n\lg \sigma)$ work and $O(\lg n\lg \sigma)$
span. In practice, the \levelWT{} algorithm shows better performance.
Compared to our previous algorithms, the \levelWT{} and \sortWT{}
algorithms can scale beyond $O(\lg\sigma)$ cores. However, both also
need to duplicate and modify the input sequence, resulting in an
increase in memory usage, requiring $O(n\lg n)$ bits of extra space.

\subsection{Problem statement}
\label{subsec:problem_statement}
The \WT{} is a versatile data structure that uses $n\lg\sigma +
o(n\lg\sigma)$ bits of space and supports several queries (such as access, rank
and select) in 
$O(\lg\sigma)$ time, for a sequence of $n$ symbols over an alphabet $\Sigma$ of size
$\sigma$. The \WT{} can
be constructed in $O(n\lg\sigma)$ time, which may be prohibitive for large
sequences. Therefore, in this work, we reduce the time complexity of the most
time-consuming operation of \WT{}, its construction, on multicore
architectures. Given a multicore machine with $p$ available cores, we propose the
design and implementation of parallel algorithms to the construction of
\WT{}. The proposed algorithms scale with $p$, achieving good practical speedups
and extra-memory usage.

\section{Multicore wavelet tree}
\label{sec:pwt}

We focus on binary wavelet trees where the symbols in $\Sigma$ are
contiguous in $[1,\sigma]$. If they are not contiguous, a bitmap is
used to remap the sequence to a contiguous
alphabet~\cite{bib:claude08}. Under these restrictions, the \WT{} is
a balanced binary tree with $\lceil \lg \sigma \rceil$ levels.  In
this section we build the representation of \WT{}s that removes the
$O(\sigma\lg n)$ bits of the topology. Hence, when we refer to a
\emph{node}, this is a conceptual node that does not exist in the
actual implementation of the data structure.

In what follows, two iterative construction algorithms are introduced
that capitalize on the idea that any level of the \WT{} can be built
independently from the others. Unlike in classical \WT{} construction,
when building a level we cannot assume that any previous step is
providing us with the correct permutation of the elements of
$S$. Instead, we compute the node at level $i$ for each symbol of the
original sequence. More formally,

\begin{proposition}\label{lem_nodecomput}
  Given a symbol $s \in S$ and a level $i$,
  $0 \leq i < l = \lceil \lg \sigma \rceil$, of a \WT{}, the node at
  which $s$ is represented at level $i$ can be computed as
  $s \gg l-i$.
\end{proposition}

In other words, if the symbols of $\Sigma$ are contiguous, then the
$i$ most significant bits of the symbol $s$ give us its corresponding
node at level $i$. In the word-RAM model with word size
$\Omega(\lg n)$, this computation takes $O(1)$ time. Since the
word-RAM model is a subset of the DYM model\footnotemark[2], the
following corollary holds:

\begin{corollary}
  The node at which a symbol $s$ is represented at level $i$ can be
  computed in $O(1)$ time.
\end{corollary}

\subsection{Per-level parallel algorithm}
\label{subsec:pwt}

Our first algorithm, called \PWT{}, is shown in Algorithm
\ref{algo:parconstruction} (the sequential version can be obtained by
replacing {\bf parfor} instructions with sequential {\bf for}
instructions). The algorithm takes as input a sequence of symbols $S$,
the length $n$ of $S$, and the length of the alphabet, $\sigma$ (see
Sect. \ref{subsec:wt}). The output is a \WT{} $WT$ that represents
$S$.  We denote the $i$th level of $WT$ as $WT[i]$,
$\forall i, 0\le i < \lceil\lg \sigma\rceil$.

\begin{algorithm}[t]
  \small
  \SetVlineSkip{-2cm}
  \SetKwInOut{Input}{Input}\SetKwInOut{Output}{Output}
  \SetKwFor{PFor}{parfor}{do}{end}
  \SetKwFunction{Kwbasb}{bitmapSetBit}
  \SetKwFunction{RankSel}{createRankSelect}
  \SetKwFunction{PrefixSum}{parPrefixSum}
  \SetKwFunction{Increment}{increment}
  \LinesNumbered
  \SetAlgoNoEnd
  \DontPrintSemicolon
  \Input{$S$, $n$, $\sigma$}
  \Output{A wavelet tree representation $WT$ of $S$}
  \BlankLine
  $WT$ is a new wavelet tree with $\lceil \lg \sigma \rceil$ levels\;
  \PFor{$i\asgn 0$ \KwTo $\lceil \lg \sigma \rceil-1$}{
    $B$ is a bitarray of size $n$\;
    $C$ is an integer array of size $2^i$\;

    \For{$j\asgn 0$ \KwTo $n-1$}{
      \Increment{$C[S[j]/2^{\lceil \lg \sigma \rceil-i}]$}\;
    }

    \PrefixSum{C}\;

    \For{$j \asgn 0$ \KwTo $n-1$}{
      \eIf{$(S[j]\ \&\ 2^{\lceil \lg \sigma \rceil-i-1}) == 1$ }{
        \Kwbasb{$B, C[S[j]/2^{\lceil \lg \sigma \rceil-i}], 1$}\;
      }
      {
        \Kwbasb{$B, C[S[j]/2^{\lceil \lg \sigma \rceil-i}], 0$}\;
      }
      
      \Increment{$C[S[j]/2^{\lceil \lg \sigma \rceil-i}]$}\;
    }

    $WT[i] \asgn $ \RankSel{B}\;
    
  }
  \Return{$WT$}\;
  \caption{Per-level parallel algorithm ({\tt pwt})}
  \label{algo:parconstruction}
\end{algorithm}
\normalsize

\begin{figure}[b]
  \centering
  \includegraphics[scale=.7]{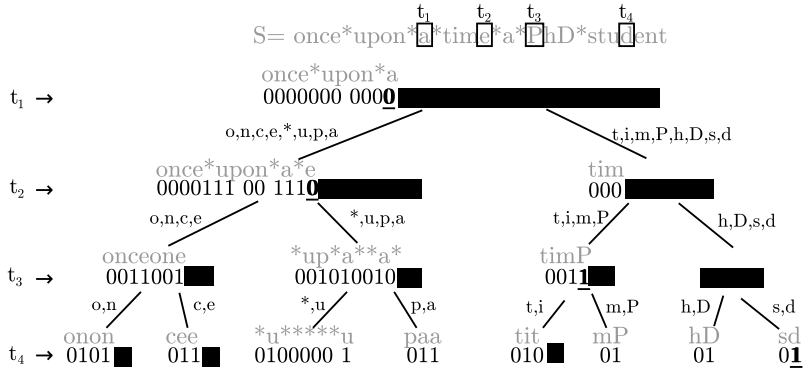}
  \caption{Snapshot of an execution of the algorithm \PWT{} for the
  sequence introduced in Figure \ref{fig:wtree}. In the
  snapshot, thread $t_1$ is writing the first bit of the symbol $S[10]=\mlq
  a\mrq$ at level $0$, thread $t_2$ is writing the second bit of $S[15]=\mlq
  e\mrq$ at level $1$, thread $t_3$ is writing the third bit of $S[19]=\mlq
  P\mrq$ at level $2$ and thread $t_4$ is writing the fourth bit of $S[26]=\mlq
  d\mrq$ at level $3$. Black areas represent bits associated to unprocessed
  symbols.}
  \label{fig:pwt}
\end{figure}

The outer loop (line 2) iterates in parallel over
$\lceil \lg \sigma \rceil$ levels. Lines 3 to 14 scan each level
performing the following tasks: the first step (lines 3 and 4)
initializes the bitmap $B$ of the $i$th level and initializes an array
of integers $C$. The array $C$ will be used to count the number of
bits in each node of the \WT{} at level $i$, using {\em counting
  sort}. The second step (lines 5 and 6) computes the size of each
node in the $i$th level performing a linear-time sweep over $S$. For
each symbol in $S$, the algorithm computes the corresponding node for
alphabet range at the current level. The expression
$S[j]/2^{\lceil \lg \sigma \rceil-i}$ in line 6 shows an equivalent
representation of the idea in Proposition~\ref{lem_nodecomput}. The
third step performs a parallel prefix sum
algorithm~\cite{Helman2001265} over the array $C$, obtaining the
offset of each node. Once the offset of the nodes is known, the
algorithm constructs the corresponding bitarray $B$, sequentially
scanning $S$ (lines 8 to 13).  For each symbol in $S$, the algorithm
computes the corresponding node and whether the symbol belongs to
either the first or second half of $\Sigma$ for that node. The
corresponding bit is set using \emph{bitmapSetBit} at position
$C[S[j]/2^{\lceil \lg \sigma \rceil-i}]$.  Line 14 creates the
rank/select structures of the bitmap $B$ of the $i$th level.

Figure \ref{fig:pwt} shows an snapshot of the execution of
  the \PWT{} for the input sequence of Figure \ref{fig:wtree}: the
  levels of the \WT{} can be constructed in different threads
  asynchronously.

The work $T_1$ of this algorithm takes $O(n\lg\sigma)$ time.
This matches the time for construction found in the literature.  Each
of the $\lg\sigma$ tasks that create the \PWT{} algorithm has a
complexity of $O(n+\sigma/p+\lg p)$, due to the scans over the input
sequence and the parallel prefix sum over the array $C$.  The work of
\PWT{} is still $T_1=O(n\lg\sigma)$.  Since all tasks have the same
complexity, assuming constant access to any position in memory, the
critical path is given by the construction of one level of the
\WT{}. That is, for $p=\infty$, $T_\infty=O(n+\lg\sigma)=O(n)$.
In the same vein, parallelism will be $T_1/T_\infty=O(\lg\sigma)$.
It follows that having $p\leq\lg\sigma$ the algorithm will obtain
optimal speedup. The overhead added for the {\bf parfor},
$O(\lg\lg\sigma)$ is negligible.
With respect to the working space, the algorithm
\PWT{} needs the space of the \WT{} and the extra space for the array
$C$, that is, a working space of $O(n\lg\sigma+\sigma\lg n)$ bits.

The main drawback of the \PWT{} algorithm is that it only scales linearly until
the number of cores equals the number of levels in the wavelet
tree. So, even if we have more cores available, the algorithm will
only use efficiently up to $\lg\sigma$ cores.  Nevertheless, this algorithm is
simple to implement, and suitable in domains where there is not possible to
use all available resources to the construction of \WT{}s.

\subsection{Domain decomposition parallel algorithm}
\label{subsec:dd}

The second algorithm that we propose makes efficient use of all
available cores. The main idea of the algorithm is to divide the input
sequence $S$ in $k=O(p/\lg(\sigma))$ segments of size $O(n/k)$ and
then apply the \PWT{} algorithm on each segment, generating
$O(\lg\sigma)$ tasks per segment and creating $k$ partial \WT{}s.
After that, the algorithm merges all the partial \WT{}s into a single
one that represents the entire input text. We call this algorithm
\DD{} because of its domain decomposition nature. This
  algorithm improves the $O(n)$ time complexity of the one introduced
  previously in \cite{Fuentes2014}.

\begin{algorithm}[t]
  \small
  \SetVlineSkip{-2cm}
  \SetKwInOut{Input}{Input}\SetKwInOut{Output}{Output}
  \SetKwFor{PFor}{parfor}{do}{end}
  \SetKwFunction{Kwbasb}{bitmapSetBit}
  \SetKwFunction{CreatePartial}{createPartialBA}
  \SetKwFunction{ParPrefixSum}{parPrefixSum}
  \SetKwFunction{Merge}{mergeBA}
  \LinesNumbered
  \SetAlgoNoEnd
  \DontPrintSemicolon
  \Input{$S$, $n$, $\sigma$, $k$}
  \Output{A wavelet tree representation $WT$ of $S$}
  \BlankLine
  
  $WT$ is a new tree with $\lceil \lg \sigma \rceil$ levels\;
  $B$ is an array of $\lceil \lg \sigma \rceil$ bitarrays of size $n$\;
  $pB$ is a bidimensional array of bitarrays of dimensions $k\times \lceil \lg \sigma \rceil$\;
  $G,L$ are tridimensional arrays of integers of dimensions $k\times \lceil \lg \sigma \rceil\times 2^{level}$\;

  \PFor{$i\asgn 0$ \KwTo $k-1$}{
    $pB[i] \asgn $ \CreatePartial{S,$\sigma$,i,n/k}
  }   
  
  \PFor{$i\asgn 0$ \KwTo $\lceil \lg \sigma \rceil-1$}{
    \ParPrefixSum{i,k}\;
  }   

  $B \asgn$ \Merge{n,$\sigma$,k,pB}\;
  
  \PFor{$i\asgn 0$ \KwTo $\lceil \lg \sigma \rceil-1$}{
    $WT[i] \asgn $ \RankSel{B[i]}\;
  }
  
  \Return{$WT$}\;
  \caption{Domain decomposition parallel algorithm ({\tt dd})}
  \label{algo:ddconstruction}
\end{algorithm}
\normalsize

\begin{function}[t]
  \small
  \SetKwInOut{Input}{Input}\SetKwInOut{Output}{Output}
  \SetKwFor{PFor}{parfor}{do}{end}
  \SetKwFunction{Kwbasb}{bitmapSetBit}
  \SetKwFunction{CreatePartial}{createPartialBitarray}
  \SetKwFunction{PrefixSum}{prefixSum}
  \SetKwFunction{Merge}{mergeWT}
  \SetKwFunction{Increment}{increment}
  \LinesNumbered
  \SetAlgoNoEnd
  \DontPrintSemicolon
  \Input{$S$, $\sigma$, $k'$, $n$}
  \Output{A bitarray representation $B$ of the $k'$th segment of $S$}
  \BlankLine

  $B$ is an array of $\lceil \lg \sigma \rceil$ bitarrays of size $n$\;
  \PFor{$i\asgn 0$ \KwTo $\lceil \lg \sigma \rceil-1$}{

    \For{$j\asgn n\times k'$ \KwTo $n\times(k'+1)-1$}{
      \Increment{$G[k'][i][S[j]/2^{\lceil \lg \sigma \rceil-i}]$}\;
    }

    \PrefixSum{G,L}\;

    \For{$j \asgn n\times k'$ \KwTo $n\times(k'+1)-1$}{

      \eIf{$(S[j]\ \&\ 2^{\lceil \lg \sigma \rceil-i-1}) == 1$ }{
        \Kwbasb{$B, G[k'][i][S[j]/2^{\lceil \lg \sigma \rceil-i}], 1$}\;
      }
      {
        \Kwbasb{$B, G[k'][i][S[j]/2^{\lceil \lg \sigma \rceil-i}], 0$}\;
      }
      \Increment{$G[k'][i][S[j]/2^{\lceil \lg \sigma \rceil-i}]$}\;
    }

  }
  \Return{$B$}\;
  \caption{createPartialBA()}
  \label{algo:createPartialBitarray}
\end{function}

\begin{function}[h]
  \small\
  \SetKwInOut{Input}{Input}\SetKwInOut{Output}{Output}
  \SetKwFor{PFor}{parfor}{do}{end}
  \SetKwFunction{Concat}{parallelBitarrayConcat}
  \SetKwFunction{RankSel}{createRankSelect}
  \LinesNumbered
  \SetAlgoNoEnd
  \DontPrintSemicolon
  \Input{$n$, $\sigma$, $k$, $pB$}
  \Output{A bitarray representation $B$ of the input sequence $S$}
  \BlankLine

  $B$ is an array of $\lceil \lg \sigma \rceil$ bitarrays of size $n$\;
  \PFor{$i\asgn 0$ \KwTo $\lceil \lg \sigma \rceil-1$}{
    \PFor{$j\asgn 0$ \KwTo $k-1$}{
      \PFor{$m\asgn j\times 2^{i}$ \KwTo $(j+1)\times 2^{i}$}{
        $dst \asgn B[i]$\tcp*[f]{Destination of the bits to be copied}\;
        $src \asgn pB[m\ mod\ k][i]$\tcp*[f]{Source of the bits to be copied}\;
        $go \asgn G[m\ mod\ k][i][m/k]$\tcp*[f]{Offset in $dst$}\;
        $lo \asgn L[m\ mod\ k][i][m/k]$\tcp*[f]{Offset in $src$}\;
        $nb \asgn L[m\ mod\ k][i][m/k+1]-L[m\ mod\ k][i][m/k]$\tcp*[f]{Number of bits}\;
        \Concat{dest,src,go,lo,nb}
      }
    }
  }
  
  \Return{$B$}\;
  \caption{mergeBA()}
  \label{algo:mergeWT}
\end{function}
\normalsize
   
The \DD{} algorithm is shown in Algorithm
\ref{algo:ddconstruction}. It takes the same input as \PWT{} with the
addition of the number of segments, $k$.  The output is a \WT{} $WT$,
which represents the input data $S$.

The first step of \DD{} (lines 1 to 4) allocates memory for the output
\WT{}, its bitarrays, $B$, the bitarrays of the partial \WT{}s, $pB$,
and two 3-dimensional arrays of numbers, $L$ and $G$, where the third
dimension changes according to the number of nodes in each
level. Arrays $L$ and $G$ store local and global offsets,
respectively.  The local offsets store the offsets of all the nodes of
the partial \WT{}s with respect to the partial \WT{} containing
them. Similarly, $G$ stores the offsets of all the nodes of the
partial \WT{}s with respect to the final \WT{}. In other words, each
entry $L[a][b][c]$ stores the position of node $c$ at level $b$ whose
parent is partial \WT{} $a$. Each entry $G[a][b][c]$ stores the
position of node $c$ at level $b$ in the partial \WT{} $a$ inside the
final \WT{}. We will treat the arrays $L$ and $G$ as global variables
to simplify the pseudocode.

The second step (lines 5 and 6) computes the partial \WT{}s of the $k$
segments in parallel. For each segment,
\ref{algo:createPartialBitarray} is called to create the partial
\WT{}. This function is similar to the one in the \PWT{} algorithm,
performing a prefix sum (line 5 in Function
\ref{algo:createPartialBitarray}) to compute the local offsets and
store them both in $G$ and $L$. We reuse the array $G$ to save memory
in the next step. Notice that the output of the function is a partial
\WT{} composed of $\lceil \lg \sigma \rceil$ bitarrays, without
rank/select structures over such bitarrays.

The third step of the \DD{} algorithm uses the local offsets stored in
$L$ to compute the global ones (lines 7 and 8). To do that, at each
level $i$, the algorithm applies a parallel prefix sum algorithm using
the $k$ local offsets of that level. The prefix sum algorithm uses the
implicit total order within the local offsets. Since each level in the
offsets is independent of the others, we can apply the
$\lceil \lg \sigma \rceil$ calls of the parallel prefix sum algorithm
in parallel.

Once we have the global offsets computed, the fourth step merges all
partial \WT{}s, in parallel. Function \ref{algo:mergeWT}
creates one parallel task for each node in the partial
\WT{}s. In each parallel task (lines 5 to 10) the function
concatenates the bitarray of the node $m/k$ of the $i$th level of the
$m\ mod\ k$ partial \WT{} into the corresponding bitarray, $B[i]$, of the
final \WT{}. Using the local and the global offsets, the function {\tt
  parallelBitarrayConcat} copies the $nb$ of $pB[i]$, starting at
position $L[m\ mod\ k][i][m/k]$ into the bitarray $B[i]$ at position
$G[m\ mod\ k][i][m/k]$. The function {\tt parallelBitarrayConcat} is
\emph{thread-safe}: the first and last machine words that compose each
bitarray are copied using atomic operations. Thus, the
  concatenated bitarrays are correct regardless of multiple concurrent
  concatenations. The last step, lines 10-11, creates the rank/select
structures for each level of the final \WT{}.

\begin{figure}
    \centering
    \begin{subfigure}[b]{\textwidth}
        \centering
        \includegraphics[scale=.65]{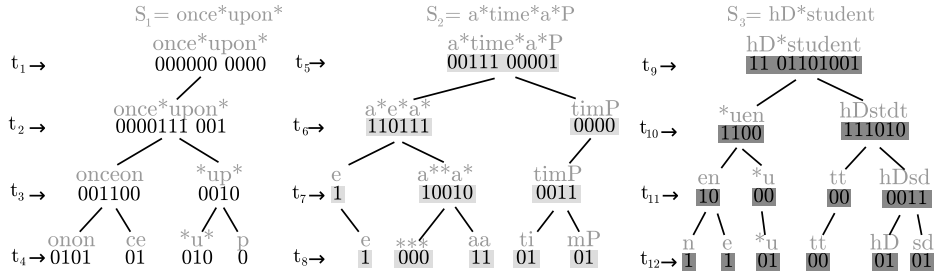}
        \caption{Snapshot of Function \ref{algo:createPartialBitarray}. The figure shows the construction
    of the partial \WT{}s after the split of the input sequence introduced in
    Figure \ref{fig:wtree} into three subsequences. To create each
    partial \WT{}, the algorithm uses the \PWT{} algorithm. These partial \WT{}s are the
    input of Function \ref{algo:mergeWT}.}
        \label{fig:partialwt}
    \end{subfigure}
    \begin{subfigure}[b]{\textwidth}
        \centering
        \includegraphics[scale=.7]{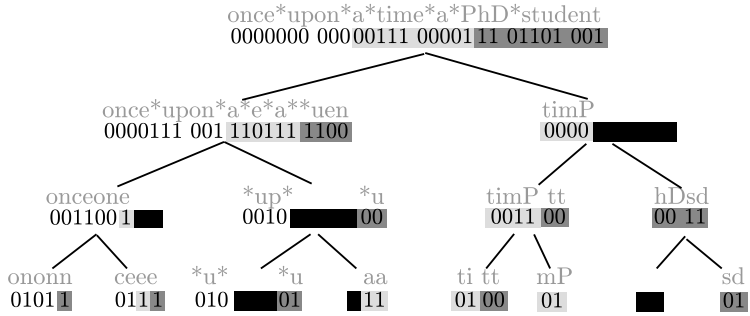}
        \caption{Snapshot of the Function \ref{algo:mergeWT}. White,
    light gray and dark gray bitarrays represent the bitarrays of first, second
    and third partial \WT{}s, respectively. The positions of the
    partial \WT{}s bitarrays are computed in advance, therefore such bitarrays can be copied
    to the final \WT{} in parallel. Black areas represent uncopied bits.}
        \label{fig:mergedd}
    \end{subfigure}
    \caption{Snapshot of an execution of the
    algorithm \DD{}. Figures \ref{fig:partialwt} and \ref{fig:mergedd} represent
    snapshots of Functions \ref{algo:createPartialBitarray}
    and \ref{algo:mergeWT}, respectively. The result of this example is the \WT{} of
  Figure \ref{fig:wtree1}.}
    \label{fig:dd}
\end{figure}

For an example of the algorithm, see Figure \ref{fig:dd}.
  Figure \ref{fig:partialwt} shows a snapshot of the function
  \ref{algo:createPartialBitarray} and Figure \ref{fig:mergedd} shows
  a snapshot of \ref{algo:mergeWT}

The \DD{} algorithm has the same asymptotic complexity as \PWT{}, with
work $T_1 = O(n\lg\sigma)$. When running on $p$ cores and dividing $S$
in $k=O(p/\lg\sigma)$ segments, the construction of the partial \WT{}s
takes $O(n\lg\sigma/p)$ time. The prefix sum takes
$O(\sigma/\lg\sigma+\lg p)$ time \cite{Helman2001265}. Merge takes
$O(n\lg\sigma/pw)$, where $w$ is the word size of that
architecture. The overhead of the {\bf parfor}s is $O(\lg
p+\lg\sigma\lg\lg\sigma)$. For $p=\infty$, the span of the construction of the
partial \WT{}s is $O(1)$, $O(\lg(k\sigma))$ for the prefix sum section
and $O(1)$ for the merge function. In the case of the merge
  function, the offsets of the bitarrays have been previously computed
  and each bit can be copied in parallel. Thus, considering $w$ as a
constant and $k=O(p/\lg\sigma)$, the span is $T_\infty=O(\lg n)$ in
all cases.

The working space needed by \DD{} is limited by the space needed for
the \WT{}, the partial \WT{}s, and local and global offsets, totalling
$O(n\lg\sigma+k\sigma\lg n)$ bits. By manipulating the value of $k$,
however, we can reduce the needed space or improve the performance of
\DD{} algorithm. If $k=1$, then space is reduced to
$O(n\lg\sigma+\sigma\lg n)$ bits, but this limits scalability to
$p<\lg\sigma$. If $k=p$, we improve the time complexity, at the cost
of $O(n\lg\sigma+p\sigma\lg n)$ bits.

\section{Experimental evaluation}
\label{sec:exps}

\begin{table}[t] \centering \small
	\begin{tabular}{rlrr} \hline & Dataset & $n$ & $\sigma$ \\
\hline 1 & rna.512MB & 536,870,912 & 4 \\ 2 & rna.1GB & 1,073,741,824
& 4 \\ 3 & rna.2GB & 2,147,483,648 & 4 \\ 4 & rna.3GB & 3,221,225,472
& 4 \\ 5 & rna.4GB & 4,294,967,296 & 4 \\ 6 & rna.5GB & 5,368,709,120
& 4 \\ 7 & rna.6GB & 6,442,450,944 & 4 \\ 8 & rna.13GB &
14,570,010,837 & 4 \\ 9 & prot & 1,184,051,855 & 27 \\ 10 & src.200MB
& 210,866,607 & 230 \\ 11 & src.98MB & 25,910,717 & 2,446,383 \\ 12 &
src.512MB & 134,217,728 & 2,446,383 \\ 13 & src.1GB & 268,435,455 &
2,446,383 \\ 14 & src.2GB & 536,870,911 & 2,446,383 \\ 15 & en.x.27 &
134,217,728 & $2^x$ \\ 16 & en.x.28 & 268,435,456 & $2^x$ \\ 17 &
en.x.29 & 536,870,912 & $2^x$ \\ 18 & en.x.30 & 1,073,741,824 & $2^x$
\\ \hline
\end{tabular}
\caption{Datasets used in the experiments. In the datasets 15-18, $x$ can take
  values in $\{4,6,8,10,12,14\}$.}
\label{table:dataSets}
\end{table}

We tested the implementation of our parallel wavelet tree construction
algorithms considering one pointer per level and without considering
the construction time of rank/select structures. We compared our
algorithms against {\sc Libcds}\footnote{We also tested a new version
of {\sc Libcds} called {\sc Libcds2}, however the former had better
running times for the construction of \WT{}s.}, {\sc Sdsl} and the fastest
algorithm in \cite{Shun2015}. Both
libraries were compiled with their default options and the -O2
optimization flag. With regards to the bitarray implementation, we use
the 5\%-extra space structure presented in \cite{GGMNwea05} (as {\sc
Libcds} does). For {\sc Sdsl} we use the {\tt bit\_vector}
implementation with settings {\tt rank\_support\_scan<1>}, {\tt
select\_support\_scan<1>} and\\{\tt select\_support\_scan<0>} to skip
construction time of rank/select structures. In our experiments, {\tt
shun} is the fastest of the three algorithms introduced in
\cite{Shun2015}, compiled also with the -O2 optimization flag.  Our
\DD{} algorithm was tested with $k=p$ privileging time performance
over memory.

\begin{table}[t] \centering \small
   \begin{tabular}{l|rrrrrrrr}\cline{2-9} \hline
\multirow{2}{*}{Datasets} & \multirow{2}{*}{\tt libcds} &
\multirow{2}{*}{\tt sdsl} & \multicolumn{2}{c}{\PWT{}} &
\multicolumn{2}{c}{\DD{}} & \multicolumn{2}{c}{{\tt shun}}\\
\cline{4-9} & & & 1 & 64 & 1 & 64 & 1 & 64 \\
\hline rna.512MB & 23.42 & 32.41 & \underline{11.83} & 7.00 & 12.65 &
\textbf{0.40} & 12.63 & 0.67\\
rna.1GB & 47.38 & 65.30 & \underline{23.89} & 16.19 & 25.30 &
\textbf{0.62} & 25.36 & 1.32\\
rna.2GB & 100.13 & 131.86 &
\underline{46.98} & 27.62 & 50.80 & \textbf{1.20} & 50.89 & 2.64\\
rna.3GB & 142.90 & 220.11 & 71.09 & 41.00 & 75.37 & \textbf{2.17} &
\underline{66.35} & 3.79\\
rna.4GB & \multicolumn{1}{c}{-} & 198.10 &
\underline{94.39} & 55.04 & 101.44 & \textbf{2.84} &
\multicolumn{1}{c}{-} & \multicolumn{1}{c}{-}\\
rna.5GB &
\multicolumn{1}{c}{-} & 329.27 & \underline{117.13} & 68.24 & 126.66 &
\textbf{3.57} & \multicolumn{1}{c}{-} & \multicolumn{1}{c}{-}\\
rna.6GB & \multicolumn{1}{c}{-} & 389.25 & \underline{141.59} & 81.80
& 152.57 & \textbf{4.35} & \multicolumn{1}{c}{-} &
\multicolumn{1}{c}{-}\\
rna.13GB & \multicolumn{1}{c}{-} & 881.41 &
\underline{314.86} & 330.44 & 333.14 & \textbf{10.75} &
\multicolumn{1}{c}{-} & \multicolumn{1}{c}{-}\\
prot & 104.40 & 142.67
& \underline{58.54} & 21.81 & 68.19 & \textbf{2.17} & 64.06 & 3.54\\
src.200MB & 24.81 & 31.41 & \underline{14.68} & 2.67 & 17.70 &
\textbf{0.52} & 16.73 & 1.06\\
src.98MB & 7.92 & 9.52 & 5.28 & 0.77 &
5.73 & 3.94 & \underline{5.07} & \textbf{0.75}\\
src.512MB & 37.77 &
49.21 & 28.94 & 5.07 & 28.98 & 5.36 & \underline{25.52} &
\textbf{3.07}\\
src.1GB & 75.48 & 99.95 & 57.99 & 8.87 & 55.36 & 9.60
& \underline{49.52} & \textbf{6.17}\\
src.2GB & 150.67 & 205.41 &
112.78 & 25.30 & 110.83 & 15.11 & \underline{98.11} & \textbf{11.77}\\
en.4.27 & 8.78 & 14.24 & \underline{5.75} & 1.82 & 6.50 &
\textbf{0.28} & 6.98 & 0.38 \\
en.4.28 & 15.82 & 28.53 &
\underline{11.44} & 3.67 & 12.88 & \textbf{0.40} & 12.34 & 0.77 \\
en.4.29 & 35.43 & 57.11 & \underline{23.01} & 7.22 & 25.51 &
\textbf{0.84} & 24.68 & 1.57 \\
en.4.30 & 70.00 & 113.88 &
\underline{46.10} & 14.40 & 51.06 & \textbf{1.63} & 55.56 & 3.06 \\
en.6.27 & 12.44 & 19.10 & \underline{7.98} & 1.78 & 9.58 &
\textbf{0.36} & 10.46 & 0.61 \\
en.6.28 & 22.65 & 38.37 &
\underline{15.92} & 3.33 & 19.35 & \textbf{0.52} & 18.38 & 1.17 \\
en.6.29 & 50.28 & 76.91 & \underline{31.78} & 7.08 & 37.90 &
\textbf{1.18} & 41.86 & 2.36 \\
en.6.30 & 99.66 & 153.72 &
\underline{63.62} & 15.90 & 76.59 & \textbf{2.20} & 83.29 & 4.68 \\
en.8.27 & 15.87 & 26.00 & \underline{11.48} & 1.87 & 13.15 &
\textbf{0.46} & 14.10 & 0.88 \\
en.8.28 & 29.06 & 52.15 &
\underline{22.86} & 3.71 & 26.52 & \textbf{0.78} & 28.28 & 1.58 \\
en.8.29 & 64.84 & 105.01 & \underline{45.79} & 7.57 & 52.53 &
\textbf{1.56} & 56.68 & 3.14 \\
en.8.30 & 128.65 & 209.54 &
\underline{91.83} & 14.65 & 105.00 & \textbf{3.13} & 113.13 & 6.26 \\
en.10.27 & 21.32 & 33.25 & 14.61 & 2.26 & \underline{13.94} & 1.66 &
17.26 & \textbf{1.39}\\
en.10.28 & 43.55 & 68.00 & 30.32 & 6.43 &
\underline{29.05} & \textbf{2.18} & 33.15 & 2.78\\ en.10.29 & 89.96 &
136.67 & 60.69 & 9.25 & \underline{58.55} & \textbf{4.59} & 67.16 &
5.67\\
en.10.30 & 183.57 & 281.53 & 123.88 & 17.70 &
\underline{119.14} & \textbf{8.93} & 214.2 & 10.77\\ en.12.27 & 24.38
& 39.09 & 17.97 & 2.52 & \underline{17.33} & 2.61 & 20.33 &
\textbf{1.64}\\
en.12.28 & 50.17 & 80.22 & 37.66 & 7.62 &
\underline{36.36} & \textbf{2.66} & 38.97 & 3.25\\
en.12.29 & 103.39 &
161.96 & 75.09 & 10.41 & \underline{72.46} & \textbf{5.73} & 128.35 &
6.71\\
en.12.30 & 211.66 & 333.32 & 150.02 & 20.33 &
\underline{145.04} & \textbf{9.66} & 259.21 & 12.99\\
en.14.27 & 27.44
& 43.61 & 21.92 & 3.10 & \underline{21.39} & 2.43 & 22.51 &
\textbf{1.84}\\
en.14.28 & 56.44 & 90.05 & 45.85 & 6.11 & 44.70 &
\textbf{2.94} & \underline{44.53} & 3.67\\
en.14.29 & 116.15 & 182.46
& 90.41 & 12.50 & \underline{88.37} & \textbf{6.97} & 91.53 & 7.79\\
en.14.30 & 238.36 & 377.77 & 184.83 & 22.31 & \underline{178.58} &
\textbf{10.50} & 302.14 & 15.98\\
\hline
 \end{tabular}
 \caption{Running times, in seconds, of the sequential algorithms and
parallel algorithms with 1 and 64 threads. The best sequential times
are underlined and the best parallel times are shown using bold
typeface. A ``-" is shown for implementations that just work for
$n<2^{32}$.}
 \label{table:time.all}
\end{table}

\subsection{Experimental setup} All algorithms were
implemented in the C programming language and compiled with GCC 4.9
(Cilk branch) using the -O2 optimization flag. The experiments were
carried out on a 4-chip (8 NUMA nodes) AMD Opteron\textsuperscript{TM}
6278 machine with 8 physical cores per NUMA node, clocking at 2.4GHz
each, with one 64KB L1 instruction cache shared by two cores, one 16KB
L1 data cache per core, a 2MB L2 cache shared between two cores, and a
6MB of L3 shared between 8 cores per NUMA node. The machine had 192GB
of DDR3 RAM, clocking at 1333MHz, with 24GB per NUMA node. Algorithms
were compared in terms of running times using the usual
high-resolution (nanosecond) C functions in {\tt <time.h>}.  Memory
usage was measured using the tools provided by {\tt malloc\_count}
\cite{malloc-count}.

The experimental trials consisted in running the algorithms on
datasets of different alphabet sizes, input sizes $n$ and number of
cores. The datasets are shown in Table \ref{table:dataSets}. We
distinguish between two types of datasets: those in which symbols are
encoded using 1 byte, and those in which symbols are encoded using 4
bytes. Datasets 1-10 in Table \ref{table:dataSets} with $\sigma \leq
256$ were encoded using 1 byte. Datasets 11-14 were encoded using 4
bytes. Datasets 15-18, that have $\sigma=2^{x}$, were encoded as follows: for $x = \{4, 6, 8\}$,
symbols were encoded with a single byte. For $x = \{10, 12, 14\}$,
symbols were encoded in four bytes. The dataset {\tt rna.13GB} is the
GenBank mRNAs of the University of California, Santa
Cruz\footnote{\url{http://hgdownload.cse.ucsc.edu/goldenPath/hg38/bigZips/xenoMrna.fa.gz}
(April, 2015)}. The rest of the {\tt rna} datasets were generated by
splitting the previous one. We also tested datasets of protein
sequences, {\tt
prot}\footnote{\url{http://pizzachili.dcc.uchile.cl/texts/protein/proteins.gz}
(April, 2015)} and source code, {\tt
src.200MB}\footnote{\url{http://pizzachili.dcc.uchile.cl/texts/code/sources.gz}
(April, 2015)}. We also built a version of the source code dataset
using words as symbols, {\tt src.98MB}. The rest of the {\tt src}
datasets were generated by concatenating the previous one up to a
maximum of 2GB. To measure the impact of varying the alphabet size, we
took the English corpus of the Pizza \& Chili
website\footnote{\url{http://pizzachili.dcc.uchile.cl/texts/nlang/english.1024MB.gz}
(March, 2013)} as a sequence of {\em words} and filtered the number of
different symbols in the dataset. The dataset had an initial alphabet
$\Sigma$ of $\sigma$=633,816 symbols. For experimentation, we
generated an alphabet $\Sigma'$ of size $2^x$, taking the top $2^x$
{\em most frequent} words in the original $\Sigma$, and then assigning
a random index to each symbol using a Marsenne Twister
\cite{Matsumoto:1998:MTE:272991.272995}, with $x \in
\{4,6,8,10,12,14\}$. To create an input sequence $S$ of $n$ symbols
for the English dataset ({\tt en}), we searched for each symbol in
$\Sigma'$ in the original English text and, when found, appended it to
$S$ until it reached the maximum possible size given $\sigma'$
($\sim$1.5GB, in the case of $\sigma'=2^{18}$), maintaining the order
of the original English text. We then either split $S$ until we
reached the target size $n=2^{27}$ or concatenated $S$ with initial
sub-sequences of itself to reach the larger sizes $2^{28}$, $2^{29}$
and $2^{30}$. We repeated each trial five times and recorded the
median time \cite{touati:hal-00764454}\footnote{In order to
be less sensitive to outliers, we use the median time instead of other
statistics. In our experiments, the \PWT{} algorithm showed a larger
deviation with respect to the number of threads than the other
algorithms. However the differences were not statistically
significant.}.

\subsection{Running times and speedup} Table
\ref{table:time.all} shows the running times of all tested
algorithms\footnote{A complete report of running times and everything
needed to replicate these results is available at
\url{www.inf.udec.cl/~josefuentes/wavelettree}}.  {\sc Libcds} and
{\tt shun} work just for $n<2^{32}$, so we cannot report running times
of these algorithms for the datasets {\tt rna.4GB}, {\tt rna.5GB},
{\tt rna.6GB} and {\tt rna.13GB}.

For each dataset, we underline the best sequential running times. We
use those values to compute speedups. The best parallel times for
$p=64$ are identified using a bold typeface. Although {\tt libcds} and
{\tt sdsl} are the state-of-the-art in sequential implementations of
\WT{}s, the best sequential running times were obtained from the
parallel implementations running on one thread. The main reason for
this is that {\sc Sdsl} implements a semi-external algorithm for
\WT~construction, involving heavy disk access, while {\sc Libcds} uses
a recursive algorithm, with known memory and executions costs.

\begin{figure} \centering
    \begin{subfigure}[b]{0.48\textwidth}
        \includegraphics[width=\textwidth,
height=0.98\textwidth]{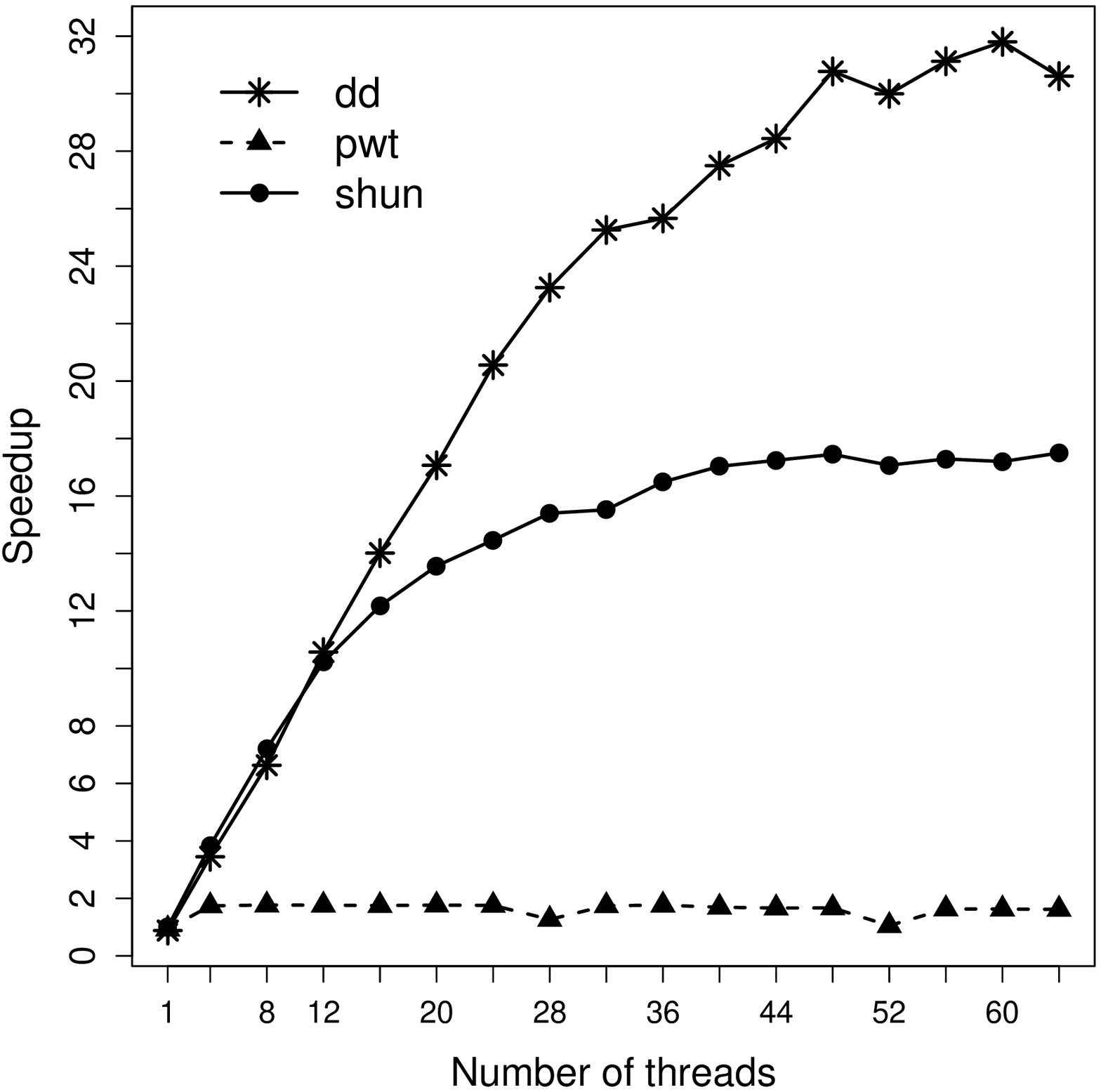}
        \caption{\emph{rna.3GB}, $n\approx2^{32}$, $\sigma$=4.}
        \label{fig:speedup.mrna}
    \end{subfigure} \quad
    \begin{subfigure}[b]{0.48\textwidth}
        \includegraphics[width=\textwidth,
height=.98\textwidth]{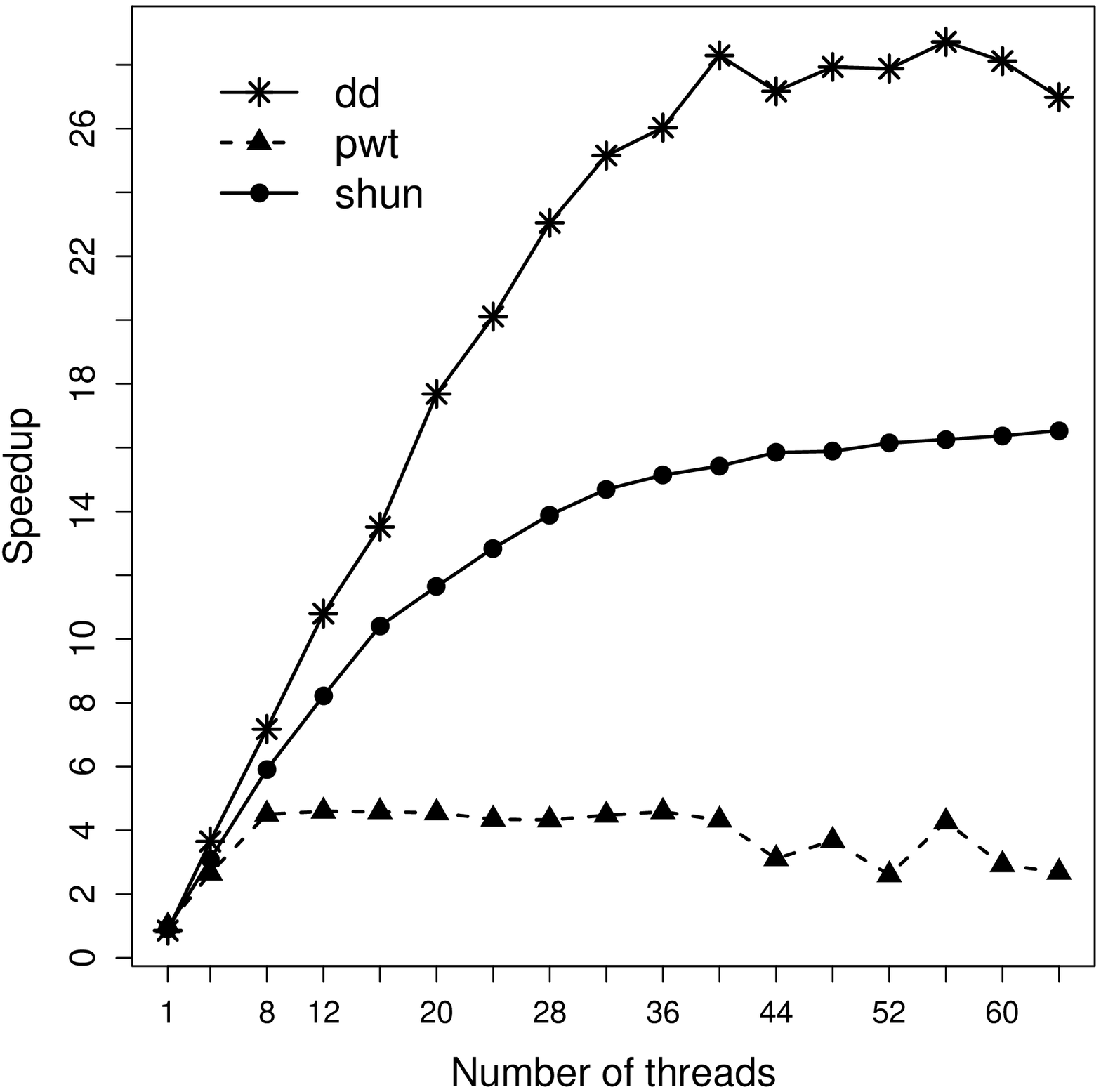}
        \caption{\emph{prot}, $n\approx2^{30}$, $\sigma$=27.}
        \label{fig:speedup.prot}
    \end{subfigure}
    \begin{subfigure}[b]{0.48\textwidth}
        \includegraphics[width=\textwidth,
height=.98\textwidth]{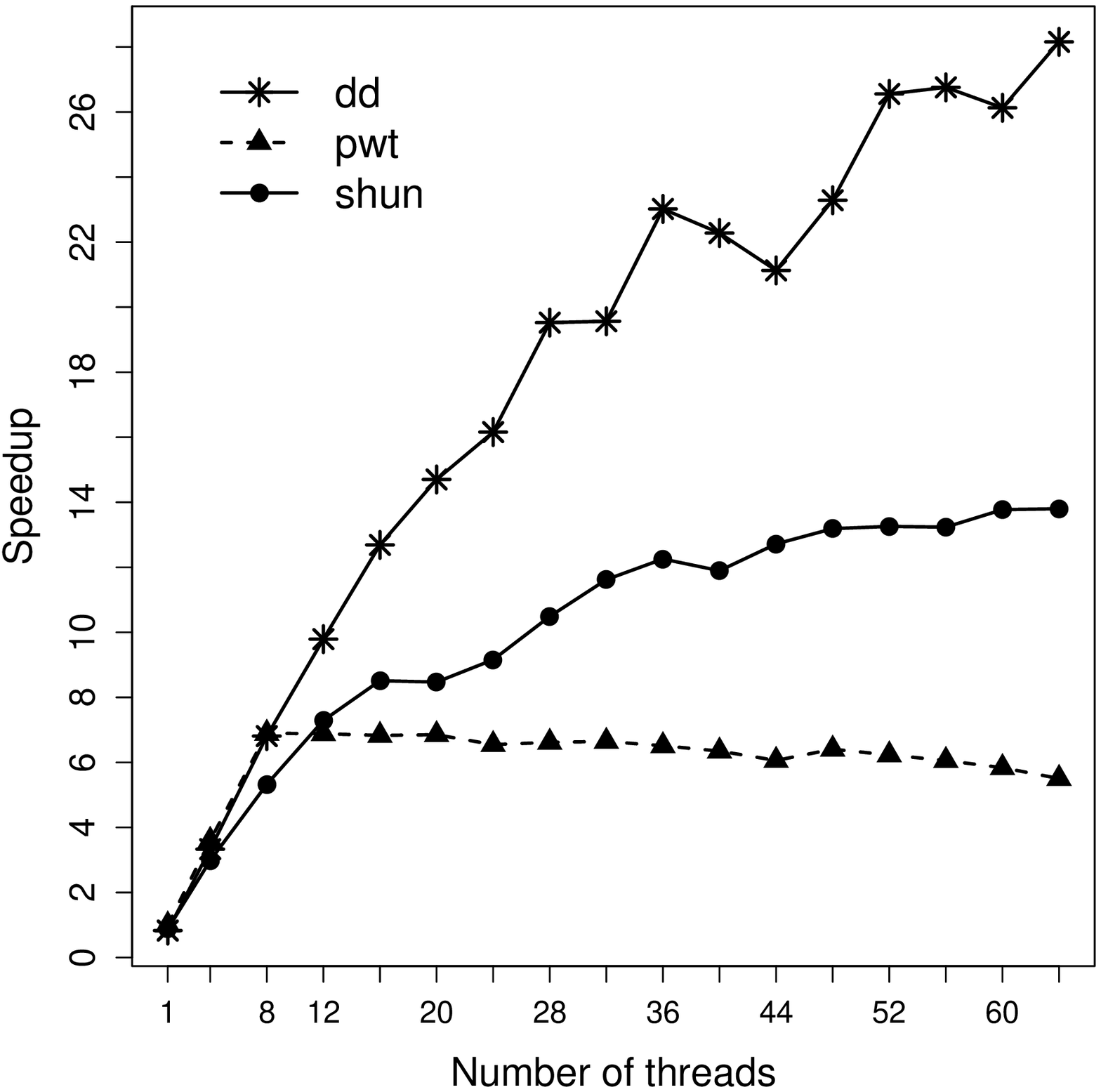}
        \caption{\emph{src.200MB}, $n\approx2^{28}$, $\sigma$=230.}
        \label{fig:speedup.src.200}
    \end{subfigure} \quad
    \begin{subfigure}[b]{0.48\textwidth}
        \includegraphics[width=\textwidth,
height=.98\textwidth]{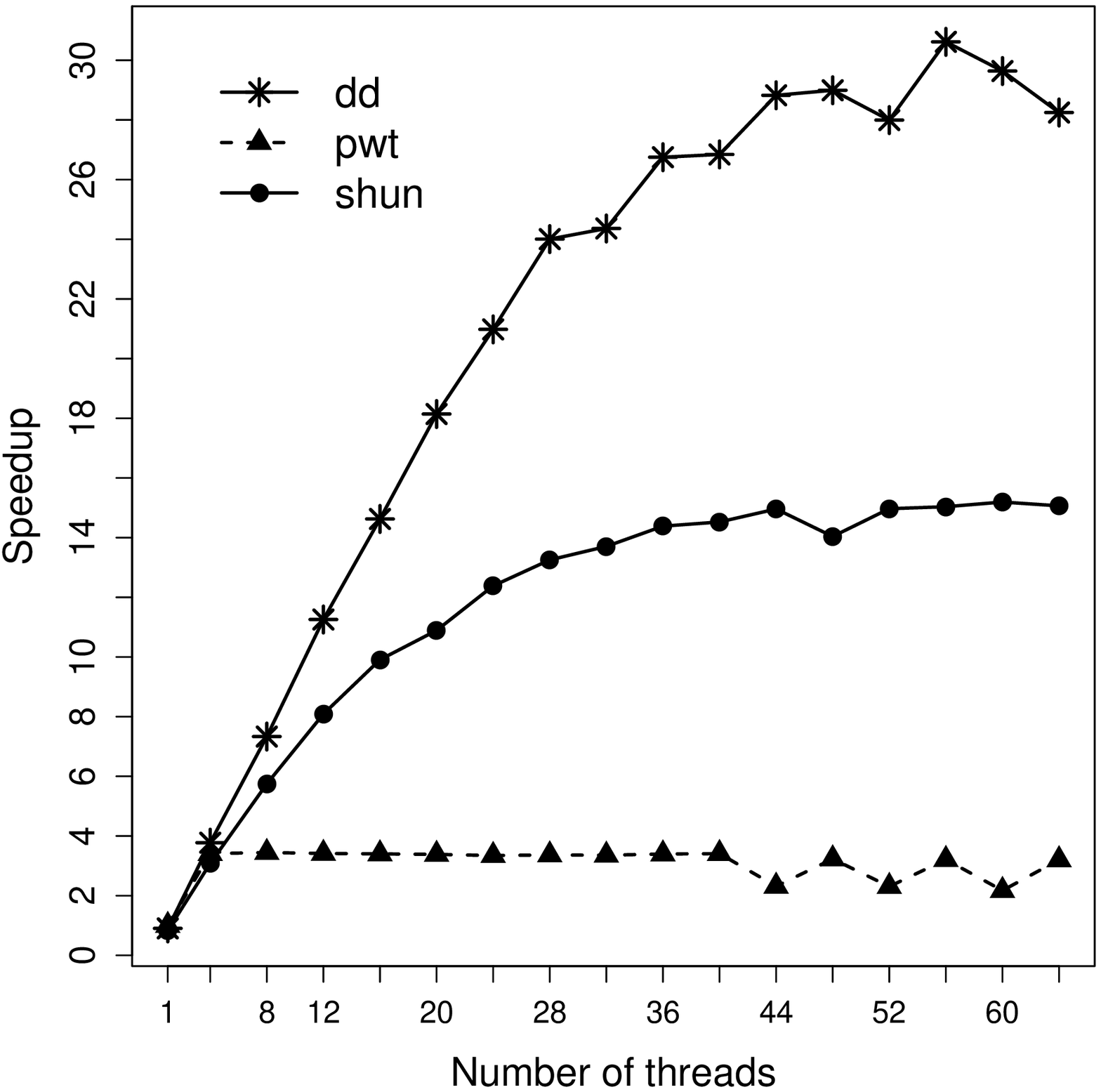}
        \caption{\emph{en.4.30}, $n=2^{30}$, $\sigma$=16.}
        \label{fig:speedup.eng.4}
    \end{subfigure}
    \begin{subfigure}[b]{0.48\textwidth}
        \includegraphics[width=\textwidth,
height=.98\textwidth]{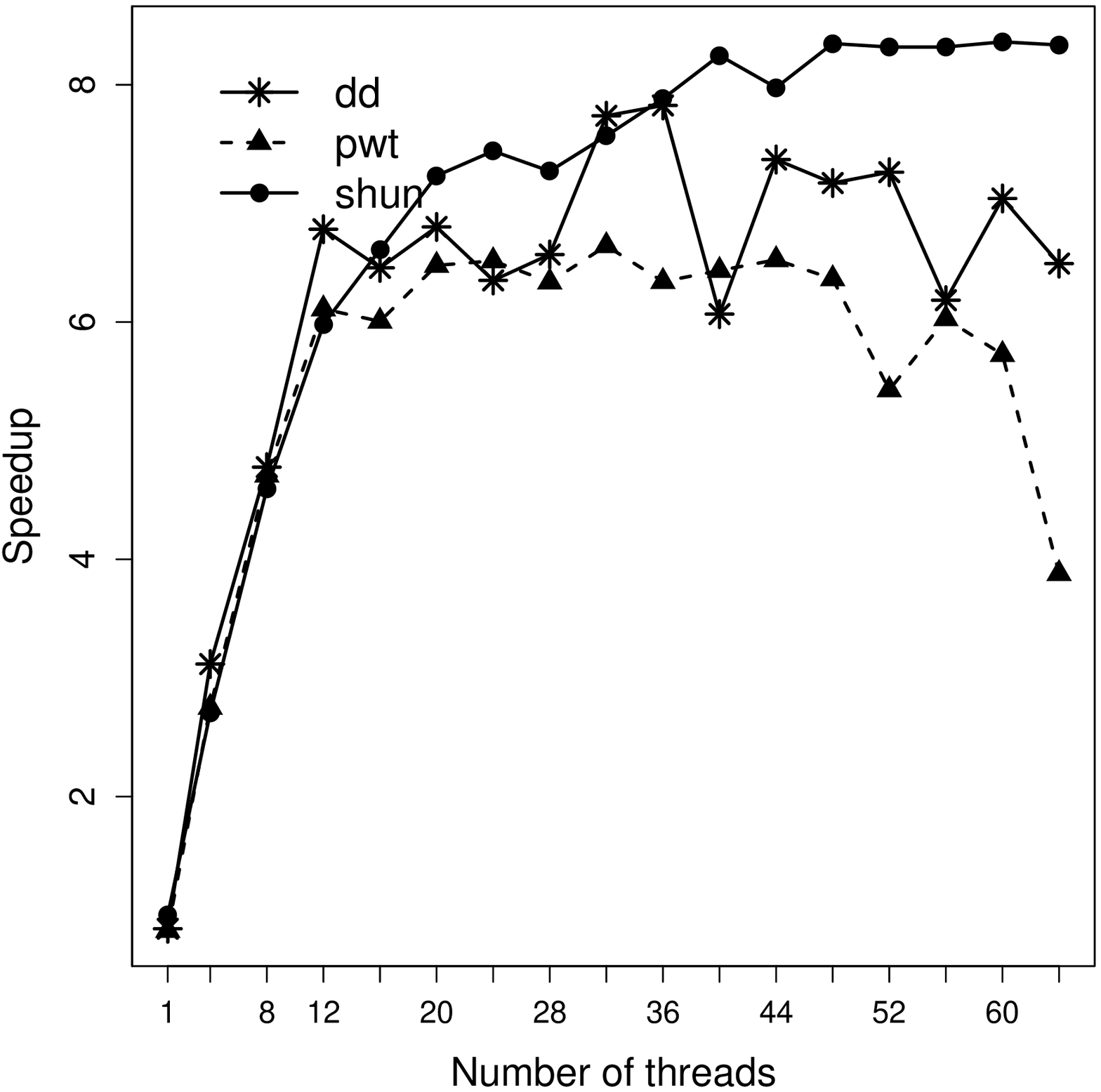}
        \caption{\emph{src.2GB}, $n\approx2^{29}$,
$\sigma\approx2^{21}$.}
        \label{fig:speedup.src.1}
    \end{subfigure} \quad
    \begin{subfigure}[b]{0.48\textwidth}
        \includegraphics[width=\textwidth,
height=.98\textwidth]{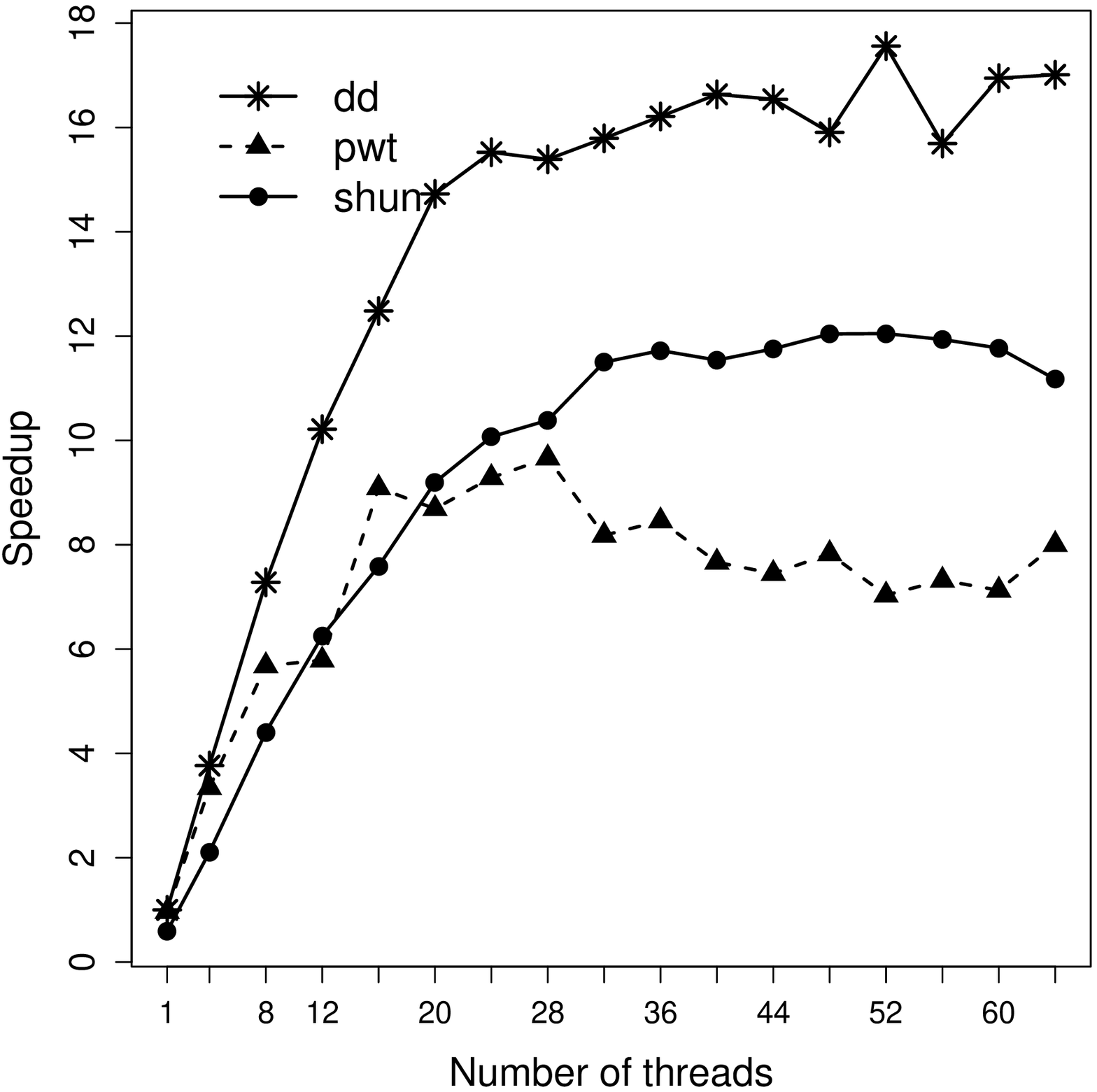}
        \caption{\emph{en.14.30}, $n=2^{30}$, $\sigma$=$2^{14}$.}
        \label{fig:speedup.eng.14}
    \end{subfigure}
    \caption{Speedup with respect to the best sequential time. The
caption of each figure indicates the name of the dataset, the input
size $n$ and the alphabet size $\sigma$.}
    \label{fig:speedup-graphs}
\end{figure}

Figure \ref{fig:speedup-graphs} shows speedups for {\tt rna.3GB}, {\tt
prot}, {\tt src.200MB}, {\tt en.4.30}, {\tt src.2GB}, {\tt en.14.30}
datasets, with the largest $n$. As expected, the \PWT{} algorithm is
competitive until $p<\lg\sigma$. Thus, for small $\sigma$ the \PWT{}
algorithm is not the best alternative as shown in Figures
\ref{fig:speedup.mrna}, \ref{fig:speedup.prot} and
\ref{fig:speedup.eng.4}. If the algorithm recruits more threads than
levels, the overhead of handling these threads increases, generating
some ``noise'' in the times obtained. The performance of \PWT{} will
be dominated also by the thread that builds more levels. For instance,
in Figure \ref{fig:speedup.eng.14} we created a \WT{} with 14
levels. In the case of one thread, that thread has to build the 14
levels. In the case of 4 threads, each has to build three levels. For
8 and 12 threads, some threads will build two levels, so those threads
dominate the running time. Finally, for the case of 16 threads, each
thread has to build at most one level. This explains the ``staircase''
effect seen for \PWT{} in Figure \ref{fig:speedup.eng.14}.

In all datasets shown in Figure \ref{fig:speedup-graphs}, except for
Figure \ref{fig:speedup.src.1}, the \DD{} algorithm has a better
speedup than both \PWT{} and {\tt shun}, especially for datasets with
small alphabets, such as {\tt rna}, {\tt prot} and {\tt en.4}. In the
case of Figure \ref{fig:speedup.src.1}, {\tt shun} has a better
speedup, because our algorithms have worse data locality, we come back
to the impact of locality of reference. It is important to remember
that although {\tt shun} has a better speedup, its memory consumption
is larger than in our algorithms, as can be seen in Section
\ref{subsec:memory_consumption}.

\subsection{Memory consumption}
\label{subsec:memory_consumption} Figure \ref{fig:mem.all} shows the
amount of memory allocated with {\it malloc} and released with {\it
free}. For all algorithms, we report the peak of memory allocation and
only considered memory allocated during construction, not memory
allocated to store the input text. The datasets are ordered
incrementally by $n$. In the case of the \DD{} algorithm, the figure
shows memory consumption for $k=1$. {\sc Libcds} and {\tt shun} use
more memory during construction time. In fact, {\tt pwt} uses up to 33
and 25 times less memory than {\sc Libcds} and {\tt shun},
respectively. Memory usage in {\tt libcds} is dominated by its
recursive nature, while {\tt shun} copies the input sequence $S$, of
$O(n\lg n)$ bits, to preserve it and to maintain permutations of it in
each iteration. Additionally, {\tt shun} uses an array of size
$O(\sigma\lg n)$ bits to maintain some values associated to the nodes
of the \WT{}, such as number of bits, the range of the alphabet, and
the offset. In our algorithms and in {\tt sdsl}, memory consumption is
dominated by the arrays which store offset values, not by the input
sequence.

The main drawback of \DD{} with respect to our own \PWT{} is its
memory consumption, since the latter increases with the alphabet size
and the number of threads.  For small alphabets, the working space of
\DD{} is almost constant. For instance, memory consumption for {\tt
rna.2GB} is 1GB, plus a small overhead for each new thread. For larger
alphabets, such as {\tt src.2GB} with $\sigma\approx 2^{22}$, the
working space increases linearly with the number of threads, using
1.46GB with 1 thread and 2.5GB with 12 threads. Fortunately, most of
the sequences used in real-world applications have an alphabet size
smaller than $2^{17}$. Such is the case of DNA sequences, the human
genome, natural language alphabets (Unicode standard),
etc.\footnote{The Unicode Consortium: \url{ http://www.unicode.org/}}.

\subsection{Other experiments}

\begin{figure} \centering
  \begin{minipage}{\textwidth} \centering
  \includegraphics[width=\linewidth]{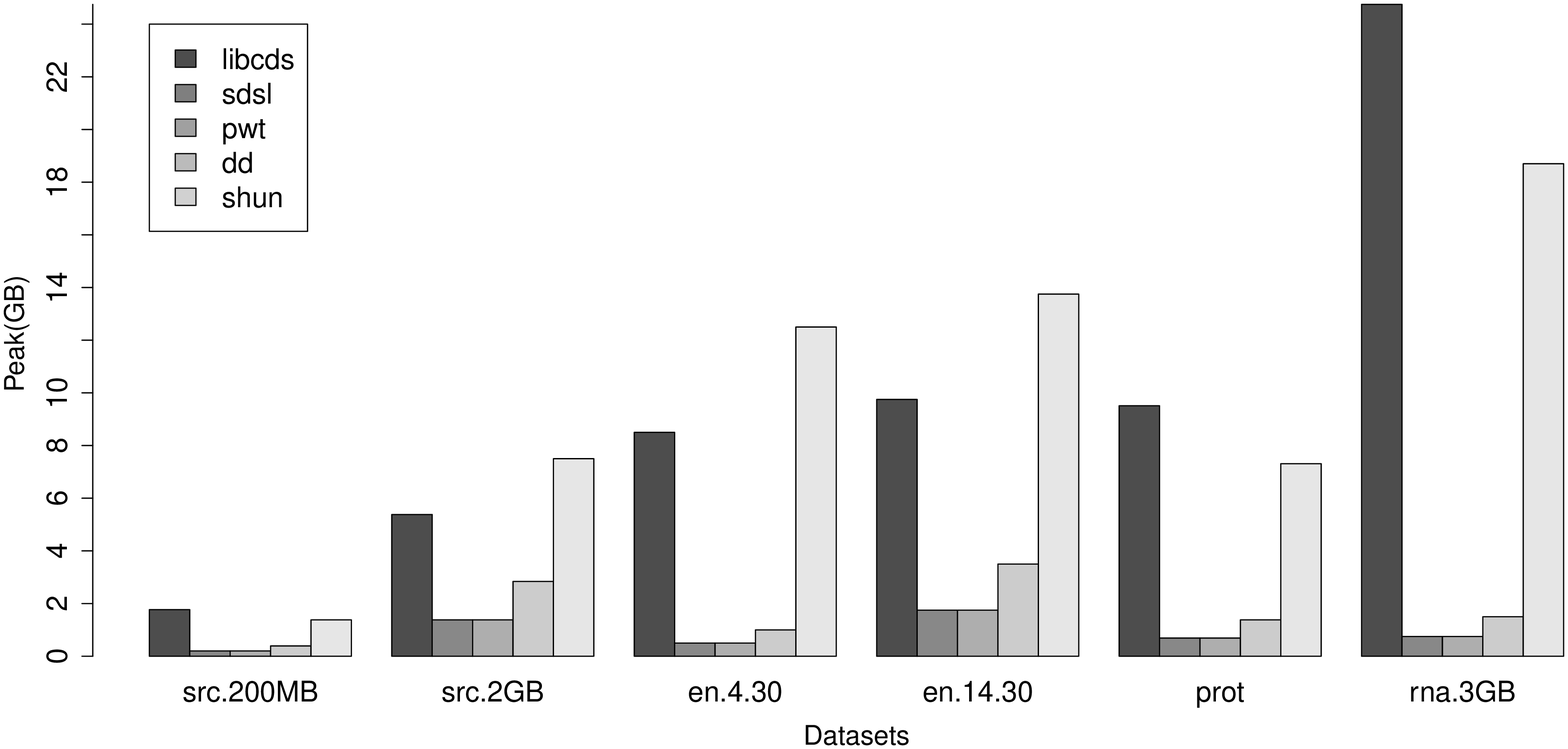}
  \caption{Memory consumption sorted by $n$.}
  \label{fig:mem.all}
\end{minipage} \centering
  \begin{minipage}{.48\textwidth} \centering
  \includegraphics[width=.8\linewidth]{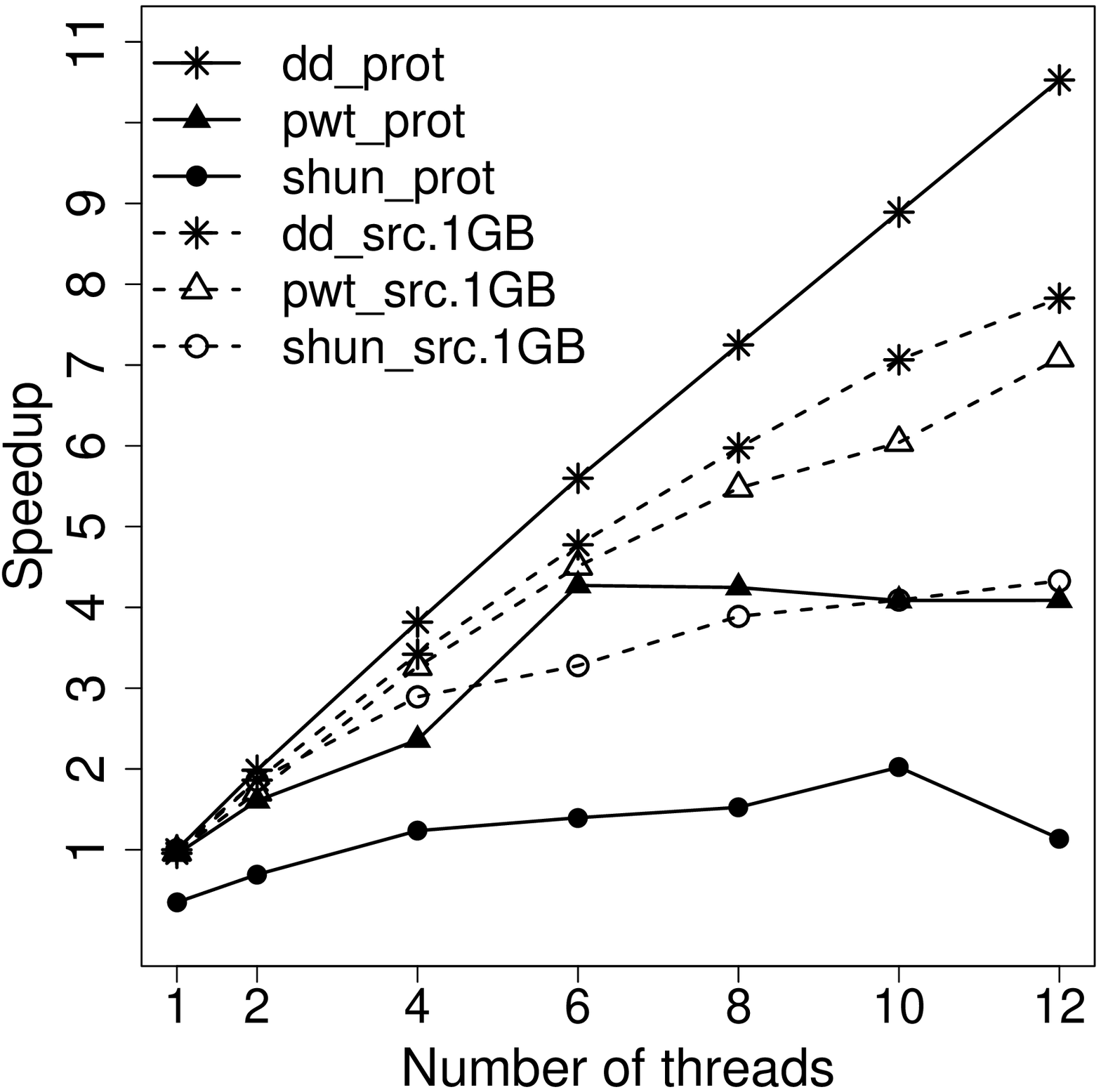}
  \caption{Running experiments in a machine with limited resources.}
  \label{fig:time.keira}
\end{minipage}
  \begin{minipage}{.48\textwidth} \centering
	 \includegraphics[width=.8\linewidth]{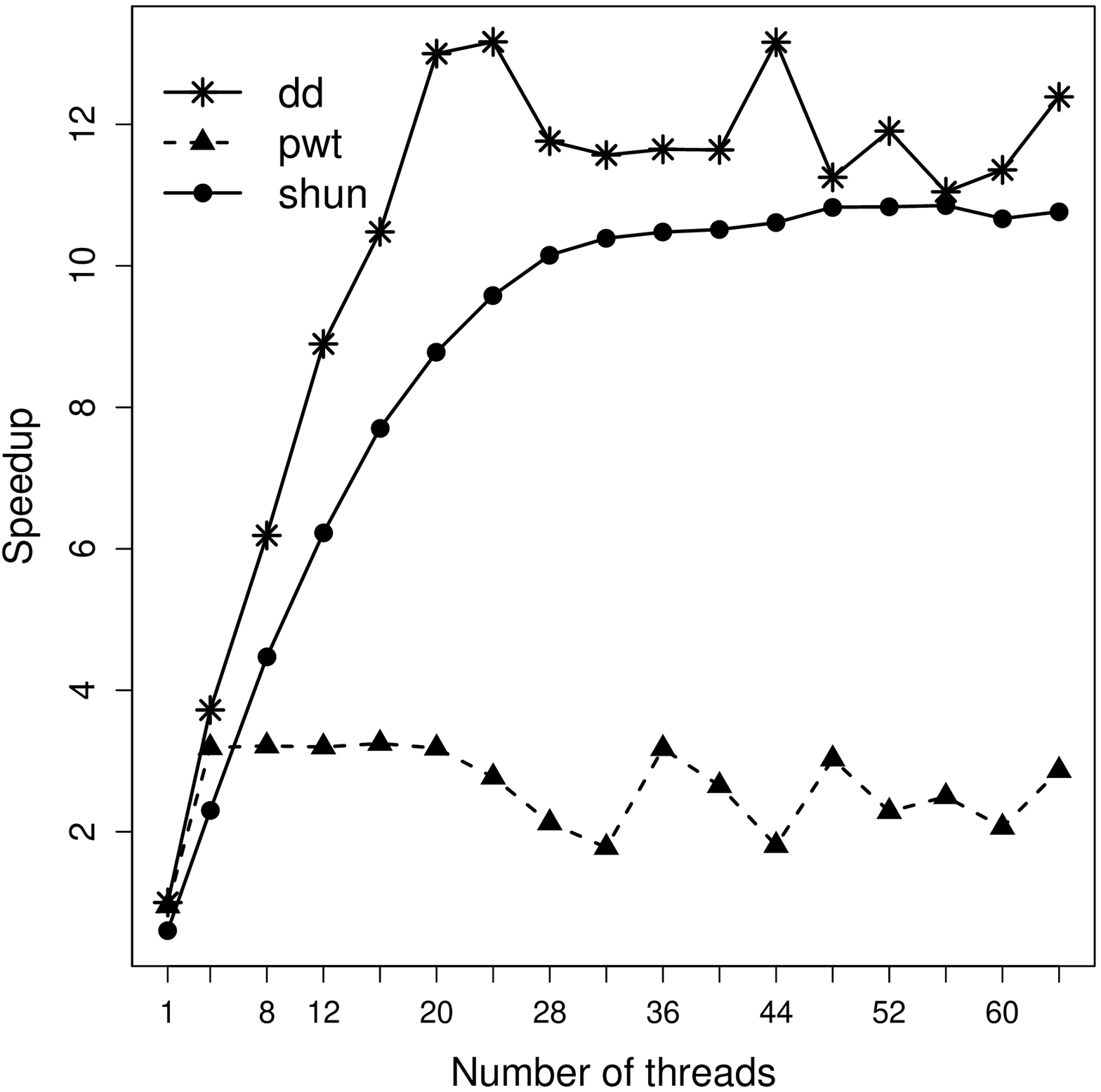}
     \caption{Speedup of the dataset \emph{en.4.30} encoding each
symbol with 4 bytes.}
     \label{fig:speedup.eng.4.4B}
\end{minipage} \centering
  \begin{minipage}{.48\textwidth} \centering
  \includegraphics[width=.8\linewidth]{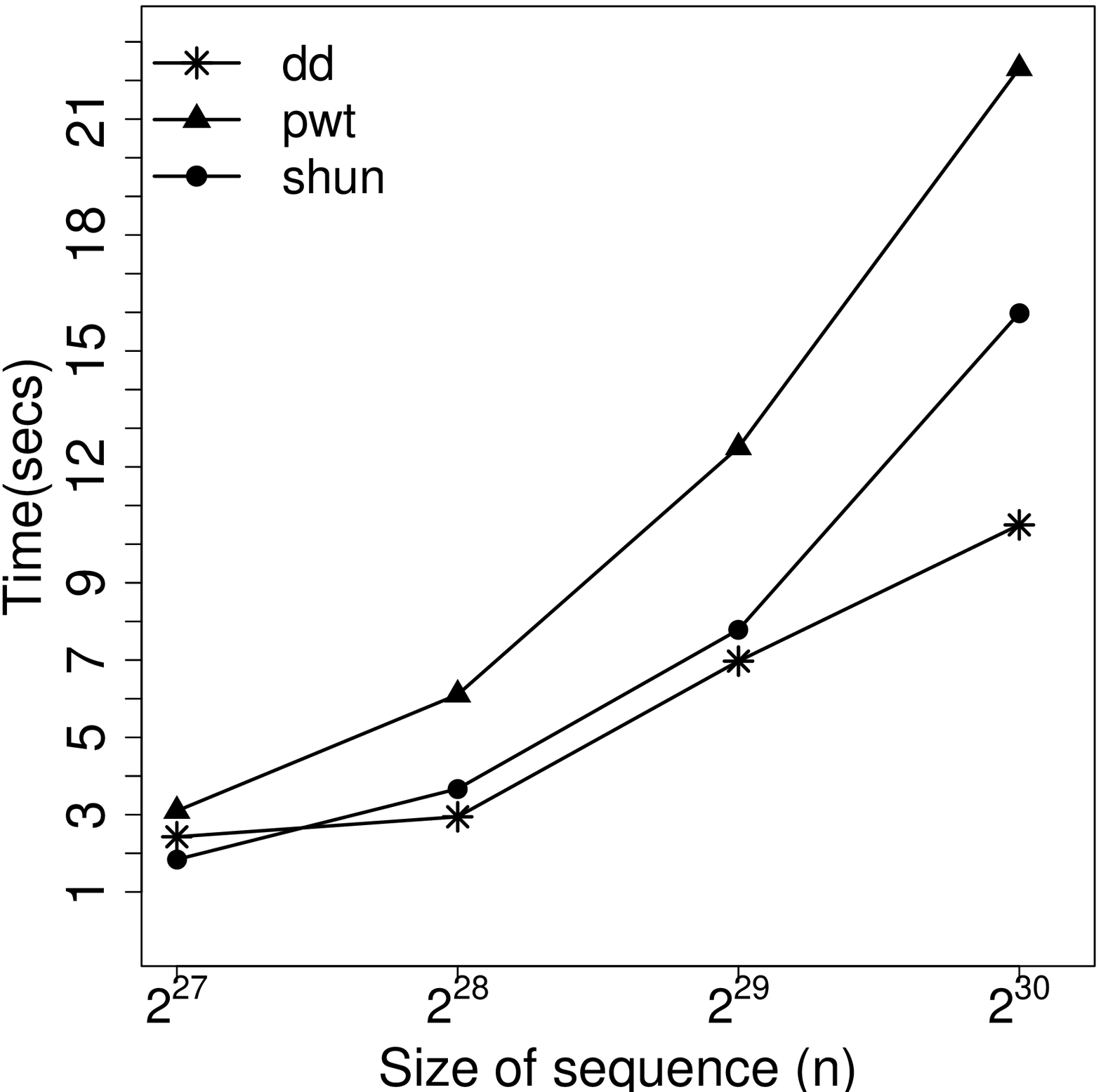}
   \caption{Time over $n$ with $\sigma=2^{14}$, 64 threads and
\emph{en.14} datasets.}
  \label{fig:time.n}
\end{minipage}
  \begin{minipage}{.48\textwidth} \centering
  \includegraphics[width=.8\linewidth]{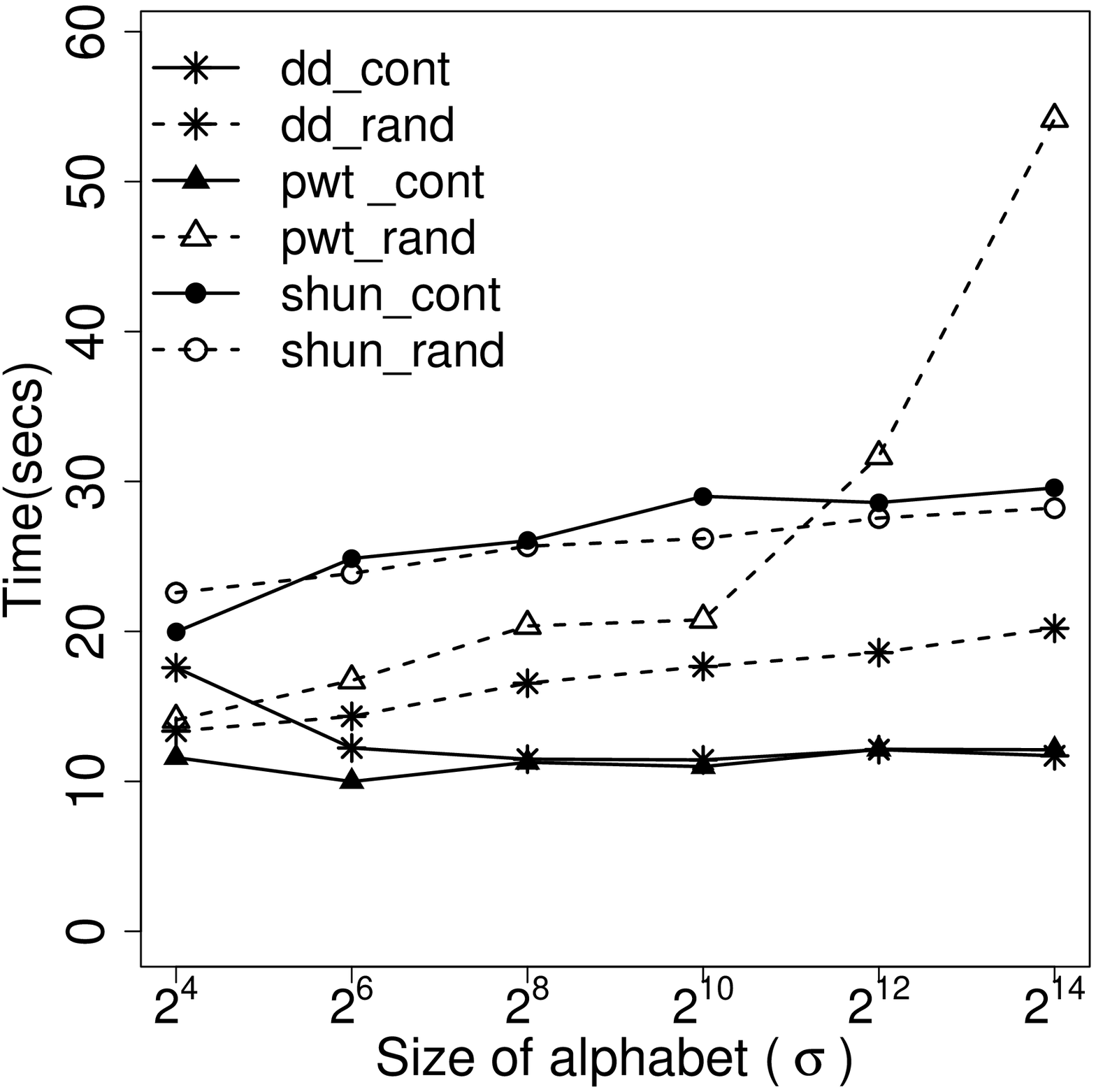}
  \caption{Time over $\sigma$ for the best and average cases with
$n=2^{30}$ and $p=\lg\sigma$ threads.}
  \label{fig:time.sigma}
\end{minipage}
\end{figure}

In order to have a better understanding of our algorithms, we
performed the following experiments:

\paragraph{Limited resources.}When memory is limited,
algorithms such as {\sc Libcds} and {\tt shun} suffer a decrement in
their performance. This is evident in Figure \ref{fig:time.keira},
where we tested the parallel algorithms with datasets {\tt prot} and
{\tt src.1GB}\footnote{The construction times of {\tt shun} with the
{\tt src.2GB} dataset exceeds one hour. To make the algorithms in the
figures comparable, we report the running times for the dataset {\tt
src.1GB}.} on a 12-core computer with 6GB of DDR3 RAM\footnote{The
computer tested is a dual-processor
Intel\textsuperscript{\textregistered}
Xeon\textsuperscript{\textregistered} CPU (E5645) with six cores per
processor, for a total of 12 physical cores running at
2.50GHz. Hyperthreading was disabled. The computer runs Linux
3.5.0-17-generic, in 64-bit mode. This machine has per-core L1 and L2
caches of sizes 32KB and 256KB respectively and 1 per-processor shared
L3 cache of 12MB, with a 5,958MB ($\sim$6GB) DDR3 RAM.}. In this new
set of experiments, the speedup of our algorithms exceeded the speedup
shown by {\tt shun}, both for datasets where we previously showed the
better performance (see Figure \ref{fig:speedup.prot}) and for
datasets where previously {\tt shun} showed better performance (see
Figure \ref{fig:speedup.src.1} and Table \ref{table:time.all}).

\paragraph{Encoding.} We observed that the encoding
of the symbols of the original sequence has a great impact in the
speedups of the construction algorithms. Figures
\ref{fig:speedup.mrna}--\ref{fig:speedup.eng.4} have speedups greater
than 27x, while there is a noticeable performance degradation in
Figure \ref{fig:speedup.src.1} and Figure
\ref{fig:speedup.eng.14}. This is due to an encoding subtlety: The
datasets used in the experiments resulting in Figures
\ref{fig:speedup.mrna}--\ref{fig:speedup.eng.4} are encoded using one
byte, while the other used four bytes. To prove the impact of the
encoding in the performance of the construction algorithms, we
repeated the experiments using a dataset that used four bytes per
symbol for $\sigma \leq 2^8$. Figures \ref{fig:speedup.eng.4} and
\ref{fig:speedup.eng.4.4B} show the influence of encoding. As
expected, the greater the memory used for encoding, the worse the
perfomance. On multicore architectures, some levels of the memory
hierarchy are shared by different cores. This increases the rate of
memory evictions. Hence, it is crucial to reduce the number of memory
transfers. Besides, in NUMA architectures, where each NUMA node has a
local RAM and the transfers between local RAMs is expensive, the
reduction of memory transfers is critical. In the case of one byte per
symbol, each memory transfer carries four times more symbols than in
the case of four bytes per symbol, effectively helping reduce memory
transfers.

\paragraph{Influence of the size of sequence.}  Figure
\ref{fig:time.n} shows that for the {\tt en.14} dataset, fixing the
number of threads to 64 and $\sigma$ to $2^{14}$, for larger $n$ the
domain decomposition algorithm behaves better in running time than the
\PWT{} algorithm and Shun's algorithm. In other words, with more cores
and enough work for each parallel task, the \DD{} algorithm should
scale appropriately.

\paragraph{Influence of the locality of reference.}
Theoretically, fixing $n$ and varying $\sigma$ with $p=\lg\sigma$
threads, the \PWT{} algorithm should show a constant-time behavior, no
matter the value of $\sigma$. However, in practice, the running times
of \PWT{} increase with the alphabet size. The reason for this
difference in theoretical and practical results is that levels closer
to the leaves in the \WT{} exhibit a weaker locality of reference. In
other words, locality of reference of the \PWT{} algorithm is
inversely proportional to $\sigma$. Additionally, the dynamic
multithreading model assumes that the cost of access to any position
in the memory is constant, but that assumption is not true in a NUMA
architecture.  In order to visualize the impact of the locality of
reference over running times, we generate two artificial datasets with
$n=2^{30}$, $\Sigma = \{1\ldots 2^{y}\}$, with $y\in\{4,6,8,10,12,
14\}$ and encoding each symbol with four bytes. The first dataset,
{\tt cont}, was created writing $n/\sigma$ times each symbol of
$\Sigma$ and then sorting the symbols according to their position in
the alphabet. The second dataset, {\tt rand}, was created in a similar
fashion, but writing symbols at random positions. The objective of the
{\tt cont} dataset is to force the best case of the \PWT{} algorithm,
where the locality of reference is higher. In contrast, the {\tt rand}
dataset forces the average case, with a low locality of reference. In
these experiments, we used the optimal number of threads of \PWT{},
that is, $p=\lg\sigma$. Besides, we allocated evenly the memory over
the NUMA nodes to ensure constant access cost to any position in the
memory\footnote{To ensure the constant access cost, we use the {\em
numactl} command with ``interleave=all'' option. The command allocates
the memory using round robin on the NUMA nodes.}.  The results are
shown in Figure \ref{fig:time.sigma}. In its average case, dashed lines,
the performance of the \PWT{} algorithm is degraded for larger
alphabets because locality of reference is low, increasing the amount
of cache misses, and thus degrading the overall performance. In the
best case, solid lines, the \PWT{} shows a practical behavior similar
to the theoretical one. Since the \DD{} algorithm implements the
\PWT{} algorithm to build each partial \WT{}, the locality of
reference impacts also on its performance. However, because the
construction of the partial \WT{}s involves sequences of size
$O(n/p)$, the impact is less than in the \PWT{} algorithm. Finally,
Shun's algorithm is insensitive to the distribution of the symbols in
the sequence.

The study of the impact of the architecture on the construction of
\WT{}s and other succinct data structures, and the improvement of the
locality of reference of our algorithms are interesting lines for
future research.

\subsection{Discussion.}

 In most cases, the domain decomposition algorithm, \DD,
  showed the best speedup. Additionally, \DD~can be adjusted either in
  favor of running time or memory consumption. \PWT{} showed good
  scalability, but up to $p<\lg \sigma$. This limitation may be
  overcome by using \PWT{} as part of \DD{}, dividing the input
  sequence in an adequate number of subsequences.

With respect to working space, \PWT~was the algorithm with
  lowest memory consumption. This is important because an algorithm
  with low memory consumption can be executed in machines with limited
  resources, can reduce cache misses due to invalidations ({\em false
    sharing}) and can therefore reduce energy consumption. Even though
  memory consumption of the \DD~algorithm increases with the number of
  subsequences, it can be controlled manipulating the number of
  segments. In the case of {\tt shun}, its memory consumption is too
  large to be competitive in machines with limited memory.

The encoding and the distribution of the symbols of the input
  sequence impact the performance of the algorithms. All the parallel
  algorithms introduced here show a better speedup for encodings that
  use less bits because there are less memory transfers. Our
  algorithms are also sensitive to the distribution of the
  symbols. When the symbols are randomly distributed, the locality of
  reference is worse in comparison with more uniform
  distributions. This gives us a hint to improve the performance of
  our algorithms in the future.

  To sum up, in general, the \DD~algorithm is the best
alternative for the construction of \WT{}s on multicore architectures,
considering both running time and memory consumption. For domains with
limited resources, \PWT, which is a building block of \DD, arises as a
good alternative on its own.

\section{Conclusions and future work}
\label{sec:conc}
Despite the vast amount of research on wavelet trees, very little has
been done to optimize them for current ubiquitous multicore
architectures. We have shown that it is possible to have practical
multicore implementations of wavelet tree construction by exploiting
information related to the levels of the \WT{}, achieving $O(\lg
n)$-time construction and good use of memory resources.

In this paper we introduced two multicore algorithms for parallel
construction of \WT{}s. Our domain decomposition algorithm, \DD{}, may
be used in any domain, but in those contexts where it is not possible
to use all available resources, our per-level algorithm, \PWT{}, may
be more suitable. We have focused on the most general representation
of a \WT{}, but some of our results may apply to other variants. For
example, it would be interesting to study how to extend our results to
compressed wavelet trees (e.g., Huffman shaped \WT{}s) and to
generalized wavelet trees (i.e., multiary wavelet trees where the fan
out of each node is increased from 2 to $O(polylog(n))$). Also, it is
interesting to explore the extension of our results to the Wavelet
Matrix~\cite{Claude12} (a different level-wise approach to avoid the
$O(\sigma \lg n)$ space overhead for the structure of the tree, which
turns out to be simpler and faster than the wavelet tree without
pointers).  Future work also involves dynamization, whereby the \WT{}
is being modified concurrently by many processes as it is queried,
though dynamic succinct data structures, even sequential ones is still
an open area of research. A further line of work involves the design
of cache-aware algorithms to construct \WT{}s, obtaining more
efficient implementations, both in time and in memory resources. In
our previous work \cite{Fuentes2014} we studied the parallelization of
some queries on \WT{}s. The parallelization of other queries is yet
another interesting future work.

For all our construction algorithms we assume that the input
  sequence $S$ fits in memory. However, we can extend our results to
  the construction of \WT{}s where the input sequence $S$ and the
  \WT{} do not fit. Following some implementation ideas of {\sc
    Sdsl}\cite{sdsl}, we can read the input sequence in buffers to
  construct partial \WT{}s for each buffer and finally merge all of
  them to obtain the final \WT. In more detail, we can extend our
  algorithms as follows:

\begin{enumerate}
  \item Read the input sequence $S$ using a buffer of size
$b$. We can use the portion of main memory that will not be used by
the \WT{} as the buffer.
  \item Create a partial \WT~without rank/select structures
taking the buffer as input. The partial \WT~can be constructed in
parallel using our \DD~algorithm with $O(b\lg\sigma/p)$ time and
$O(1)$ span. (We could also use the \PWT~if the available memory is
scarce). The starting position of each node in the partial \WT~is
stored in a bidimensional array $L$.
  \item Repeat steps 1 and 2 until the complete input
sequence is read.
  \item After the complete input sequence is read, we compute
the final position of the nodes of all the partial \WT{}s. These
positions are computed performing a parallel prefix
sum\cite{Helman2001265} over the values of the arrays $L$'s, similar
to the \DD~algorithm. It takes $O(b\sigma/p+\lg p)$ time and $O(\lg
(b\sigma))$ span.
  \item The final \WT~is constructed using Function
\ref{algo:mergeWT} with $O(n\lg\sigma/pw)$ and $O(1)$ span, where $w$
is the word size of the architecture.
\end{enumerate} The extension takes
  $O(n\lg\sigma/p+b\sigma/p+\lg p)$ time and $O(n/b+\lg(b\sigma))$
  span. Notice that this idea is similar to the \DD~algorithm and it
  can be applied on multiple levels.  For example, it can be used on
  distributed architectures, where the buffers are processed by
  different machines, and one machine merges all the partial
  \WT{}s. Additionally, observe that we can use the entire main memory
  as the buffer, storing the partial \WT{}s and the $L$ arrays on disk
  each time we finish the processing of a buffer.  We leave the
  implementation and empirical evaluation of these ideas as future
  work.

It has become evident that architecture has become relevant again. It
is nowadays difficult to find single-core computers.  Therefore, it
seems like a waste of resources to stick to sequential algorithms. We
believe one natural way to improve performance of important data
structures, such as wavelet trees, is to squeeze every drop of
parallelism of modern multicore machines.



\bibliographystyle{spmpsci}
\bibliography{wt}

\end{document}